\documentstyle[sprocl]{article}
\def\Journal#1#2#3#4{{#1} {\bf #2}, #3 (#4)}
\def\NCA{\em Nuovo Cimento}
\def\NPB{{\em Nucl. Phys.} B}

\def\PRL{\em Phys. Rev. Lett.}
\def\PR{\em Phys. Rev.}
\def\PRB{{\em Phys. Rev.} B}
\def\AdP{{\em Adv. in  Phys.} }
\def\AP{{\em Ann. Phys.} }
\def\HPA{{\em Helv. Phys. Acta} }
\def\IJMB{{\em Int. Jour. Mod. Phys.} B}
\def\CMP{{\em Comm. Math. Phys.} }
\def\MPL{{\em Mod. Phys. Lett.} A}
\begin{document}
\title{FERMIONIC CHERN-SIMONS FIELD THEORY FOR THE FRACTIONAL HALL EFFECT}
\author{ ANA LOPEZ}
\address{Department of Physics, Theoretical Physics, Oxford
University, \\ 1 Keble Rd., Oxford OX1 3NP,  United Kingdom}
\author{ EDUARDO FRADKIN}
\address{Department of Physics, University of Illinois at
Urbana-Champaign, \\1110 W. Green St., Urbana, Ill 61801, USA}

\maketitle
\abstracts{
We review the fermionic Chern-Simons field theory for the Fractional
Quantum Hall Effect (FQHE).
We show that in this field theoretic approach to the problem of
interacting electrons moving in a plane in the presence of an
external magnetic field, the FQHE states
appear naturally as the semiclassical states of the theory. In this
framework, the FQHE states are the unique ground states of a system of
electrons on a fixed  geometry. The excitation spectrum is fully gapped and
these states can be viewed as infrared stable fixed points of the system.
It is shown that the long distance, low energy properties of the
system are described exactly by this theory. It is further shown that, in
this limit, the actual ground state wave function
has the Laughlin form. We also discuss the application of this theory to the
problem of the FQHE in bilayers and in unpolarized single layer systems.}

\section { Introduction}

The fermionic Chern-Simons field theory for the FQHE \cite{lopez1}
was motivated by Jain's observation that the Laughlin wave function
could be reinterpreted as the wave function of charge-flux
composites filling up a lowest Landau level, ({\it i.~e.\/}, a
Vandermonde determinant), with each unit of charge being bound
to an even number of flux quanta.
Following Jain, we will refer to these charge-flux composite particles, as
{\it composite fermions} \cite{jain1,jain2}
In Jain's picture, the electrons ``nucleate" flux to screen enough of
the external  magnetic field, so that the ``composite fermions"
exactly fill an integer number of Landau levels associated with the
unscreened part of the field. In this formulation, the FQHE is an Integer
Quantum Hall Effect (IQHE) of the bound states.

The starting point of Jain's approach was the Laughlin wave function.
In 1989, he proposed to write it  in the suggestive factorized form
\begin{equation}
\Psi_{1\over m}(z_1, \dots, z_N)=\prod_{i<j}(z_i-z_j)^{m-1}   \;\;
\chi_1(z_1,\dots,z_N)
\label{eq:c2}
\end{equation}
where $\chi_1$ is the wave function for a completely filled lowest Landau
level
\begin{equation}
\chi_1(z_1,\dots,z_N)=\prod_{i<j}(z_i-z_j) \;\; \exp(-\sum_{i=1}^N {\vert
z_i \vert^2 \over 4 \ell^2}).
\label{eq:c3}
\end{equation}
The phases associated with the first factor in Eq.~\ref{eq:c2} represent an
even number ($m-1$) of fluxes that are attached to each coordinate
$z_i$ where
an electron is present. It is a crucial feature of this picture that
the electrons bind to an {\it even} number of flux quanta and, in this way,
they retain their fermion character.
This observation gave rise to the picture of the FQHE as a
ground state of ``composite fermions", where a composite fermion
is an electron bound to an even number of fluxes.
In this formulation, the main role of
the long range correlations is to make it possible for the electrons to
``nucleate" flux, and the FQHE can be interpreted as an Integer
Quantum Hall Effect of the bound states \cite{jain1,jain2}.

The question that naturally arises here is how does a system of electrons in
an external
magnetic field manage to turn the correlations, which result from the
electron-electron interactions, into ``nucleated fluxes".
In this Chapter we show that the key to
the answer is the Chern-Simons gauge theory.
In 1982 Wilczek's observed \cite{wilczek1} that a particle current coupled to
a Chern-Simons (CS) gauge field produced states with fractional statistics
through the binding of particles to fluxes. Thus, if we are to get the
Laughlin wave function by attaching ($m-1$) fluxes to each
electron, as suggested by Jain, it is natural to guess that the
``right" theory must contain fermions (electrons) coupled to Chern-Simons
gauge fields with an appropriate value of the Chern-Simons coupling
constant $\theta$. What is less clear is the origin of such a
Chern-Simons gauge field for the problem of interacting electrons in a
magnetic field. In Quantum Field Theory, where the CS gauge
theory was first introduced \cite{jackiw,jackiw1}, the Chern-Simons
term in the action
originates from the Parity Anomaly of relativistic fermions in 2+1 dimensions.
Clearly, the electrons which live in
quasi two-dimensional electron gases, are not
relativistic. Also the real three dimensional system one is dealing with,
neither breaks Parity by itself nor as a result of the presence of the
device. The Chern-Simons gauge field must then be a result of the dynamics of
interacting two dimensional electrons in the presence of an external magnetic
field. For our purposes,
the most important feature of the Chern-Simons term is not its relativistic
invariance, but the fact that it is the only local gauge invariant theory
which yields bound states of particles and fluxes.

In this Chapter we derive a field theory for the FQHE in the fermion
language in
which the Chern-Simons gauge field appears explicitly in the problem of
interacting {\it electrons} in a
magnetic field. We do so by first considering a theory in which the
electrons, in addition to their mutual interaction, are coupled to both an
external electromagnetic field and a Chern-Simons gauge field. In
doing so, we use the fact that
if the coefficient of the Chern-Simons
term is chosen in such a way that an {\it even} number of flux quanta get
attached to
each electron, all the physical amplitudes calculated in this theory
are identical to the amplitudes calculated in the standard theory, in
which the CS field is absent (see reference \cite{lopez1}). We further
show that the Laughlin state is the {\it semiclassical} approximation
of this theory. In the ``classical" ({\it i.~e.\/}, mean-field)
approximation we get a picture very close to the one proposed by Jain.
This theory, not only explains where do the fluxes come from, but
also allows for the systematic calculation of corrections around this
state in a manner similar to the semiclassical  approximation in
quantum mechanics and to the Random Phase Approximation (RPA)
in many-body theory.
A conceptually important feature of the Fermionic Chern-Simons theory
is that its microscopic fields are  composite fermions. Hence,
at the level of mean field theory, it describes an Integer
Quantum Hall (IQH) state of composite fermions. Nevertheless, quantum
fluctuations about this mean field state change this picture in a
qualitatively fundamental way. They  not only give the correct FQHE
but also change the charge and statistics of the quasiparticles
which become fractionally charged anyons, in agreement with the
predictions of Laughlin's theory. The reason behind this behavior is the
fact that the Chern-Simons fields require a local relation between
charge and flux. This relation is replaced at the level of mean field
theory by an average. The fluctuations correct this effect, forcing
charge fluctuations to be accompanied by the necessary flux
fluctuations. At long wavelengths, the leading gaussian fluctuations
enforce this flux-charge relation exactly, leading to the exact values
of the Hall conductance, charge and statistics of the excitations as
well as to the exact saturation of the sum rules. However, at shorter
distances and higher energies, it is necessary to go beyond gaussian
fluctuations to account for the correct correlations.

This Chapter is organized as follows. In Section \ref{sec:csfqhe}
we derive the Fermionic Chern-Simons theory of the FQHE for a single layer
fully polarized two dimensional electron gas (2DEG).
In Section \ref{sec:gaflu} we discuss the electromagnetic response
functions and the spectrum of collective modes.
In Section \ref{sec:csbil} we generalize this theory to the problem of
the FQHE in bilayers and partially polarized and spin singlet FQH states in
single layers.
In Section \ref{sec:wave} we use the Fermion Chern-Simons theory to
derive the asymptotic long distance form of the wave functions of the FQH
states. Finally, in  Section \ref{sec:conc} we present a summary of 
our results.

\section{The Fermionic Chern-Simons Theory for the Fractional Quantum
Hall Effect}
\label{sec:csfqhe}

Consider a system of $N$ electrons moving on a plane in the presence of an
external uniform magnetic field $B$ perpendicular to the plane. The electrons
will be assumed to have an interparticle interaction governed by a pair
potential $V(\vert {\vec r} \vert )$, for two electrons separated a distance
$\vert {\vec r} \vert$ on the plane. In this section we will
assume that the magnetic field is sufficiently large
so that the system is completely polarized and thus we will can ignore
the spin degrees of freedom. In Section \ref{sec:csbil} we will consider
polarization effects.
The eigenstates $\Psi({\vec x}_1,\dots,{\vec
x}_N)$ are eigenfunctions of the (first quantized) Hamiltonian ${\hat H}$
\begin{equation}
{\hat H}=\sum_{i=1}^N \lbrace {1\over 2M} \left( {\vec p}_j-{e\over
c}{\vec
A}_j({\vec x}_j) \right)^2+ eA_0({\vec x}_j) \rbrace + \sum_{i<j} V(\vert
{\vec x}_i-{\vec x}_j \vert)
\label{eq:c4}
\end{equation}
where we have included the coupling to both the electromagnetic vector
potential $\vec A$ and scalar potential $A_0$. Hence, we are dealing
with $N$ spinless fermions of charge $-e$ and mass $M$.

Our goal is to show that this
system is {\it equivalent} to the same system but coupled to an additional
{\it statistical} vector potential $a_{\mu}$ ($\mu=0,1,2$) whose
dynamics is determined by the Chern-Simons action, $S_{\rm CS}$
\begin{equation}
S_{\rm CS}=\int d^3x\;{\theta\over 4}\epsilon_{\mu \nu \lambda} a^{\mu}
{\cal F}^{\nu \lambda}.
\label{eq:c5}
\end{equation}
for a suitably chosen value of $\theta$. In Eq.~\ref{eq:c5}
$x_0,x_1 \;{\rm and}\; x_2$ represent the time and the space coordinates of
the electrons respectively, and ${\cal F}^{\nu \lambda}$ is the field tensor
for the statistical gauge field
\begin{equation}
{\cal F}^{\nu \lambda}=\partial^{\nu} a^{\lambda}-\partial^{\lambda}a^{\nu}.
\label{eq:c6}
\end{equation}
The equivalent theory has a Hamiltonian ${\hat H}'$ which is identical to
${\hat H}$, given in Eq.~\ref{eq:c4}, except for the fact that
the electrons are also coupled to the statistical vector potential
$a_{\mu}$. This is accomplished simply by setting
\begin{equation}
{e\over c}{\vec A} \to {e\over c} {\vec A}+{\vec a};\qquad
eA_0 \to eA_0+a_0.
\label{eq:c7}
\end{equation}
The Chern-Simons gauge theory is a {\it topological} field theory.
This means that the expectation values of the observables in this theory
do not depend on the distance ({\it i.~e.~}, the metric) between the
points of space-time on which they act, but only on topological
properties. Being a gauge theory, its observables are path ordered exponentials
of the circulation of gauge fields on closed loops (Wilson loops) an
their expectation values are functions of the linking numbers of these
loops \cite{witten}. Since it is a topological theory, the action of the
Chern-Simons gauge fields alone does not involve any energy scales. In
fact, it is easy to see that the Hamiltonian is actually equal to zero
(in the abscence of coupling to matter). We will now follow standard
arguments \cite{witten} and show that the content of the Chern-Simons action
reduces to a constraint and to a set of commutation relations for the gauge
fields.
Indeed, let us expand the Chern-Simons action, coupled to an external matter
current
$j_\mu(x)$, in components to read
\begin{equation}
S_{\rm CS}=\int d^3x\; \left\{[-j_0({x})+\theta {\cal B}( x)]a_0(x)+
\theta a_2(x) \partial_0 a_1(x) + {\vec a}(x) \cdot {\vec j}(x)\right\}
\label{eq:chern}
\end{equation}
where $j_o({\vec x})$ is the particle density, ${\vec j}({\vec x})$
is the matter current and
$\cal B$ is the statistical flux. By direct inspection of
Eq.~\ref{eq:chern}, we
see that the time component of the field $a_0(x)$ acts like a Lagrange
multiplier field which imposes the constraint
\begin{equation}
j_0({\vec x})=\theta {\cal B}(\vec x).
\label{eq:c8}
\end{equation}
which is the analog of Gauss' law for this theory.
At the quantum level, Eq.~\ref{eq:c8} is an {\it operator constraint}
which selects the physical space of states {\it i.~e.~\/}, the gauge
invariant states are made of charge-flux composites. The second term of the
Chern-Simons action Eq.~\ref{eq:chern} implies that the momentum
canonically conjugate to the field $a_1$ is just $\theta a_2$ and that,
quantum mechanically they have to obey the equal-time commutation
relations
\begin{equation}
\lbrack a_1 \left( {\vec x} \right) ,a_2 \left( {\vec x}'
\right) \rbrack ={i\over \theta} \delta^{(2)}({\vec x}-{\vec x}').
\label{eq:c9}
\end{equation}
Finally, the last term in the action of Eq.~\ref{eq:chern}, says that
the Hamiltonian is just
\begin{equation}
H=-\int d^2x \; {\vec j}(\vec x)\cdot {\vec a}(\vec x)
\label{eq:H1}
\end{equation}
which vanishes in the abscence of matter currents.

Thus, for {\it arbitrary} values of the Chern-Simons coupling constant
$\theta$, the physical states are charge-flux composites: every
particle of charge $1$ carries a gauge flux equal to $1/\theta$.
Intuitively we expect that the wave functions for these composite
particles should exhibit an Aharonov-Bohm effect which can be regarded
as a change of statistics \cite{wilczek,myrheim}.
In Chern-Simons theory this {\it statistical transmutation} is seen in
the following way. Imagine that we have $N$ particles which are
sufficiently heavy so
that their Feynman path integrals are dominated by reasonably smooth,
well separated paths in $2+1$-dimensional space-time. Consider for
simplicity the case of two particles and let $\Gamma_1$ and $\Gamma_2$
be their respective paths. Let us denote by $\Gamma=\Gamma_1 \bigcup
\Gamma_2$
their joint paths. Let us further define a current $j_\mu(x)$ which
carries one unit of charge and with support on $\Gamma$.
A typical amplitude for this $2$-particle system is given
by the Wilson loop expectation value \cite{witten,polyakov}
\begin{equation}
\langle e^{i\oint_{\Gamma} \; dx_\mu \; a^\mu(x)}\rangle_{\rm CS}
=
\langle e^{i\int d^3x \; j_\mu(x) a^\mu(x)}\rangle_{\rm CS}
=
e^{{\frac{i}{2\theta}} \nu_L[\Gamma]}
\label{eq:wl}
\end{equation}
where the expectation values are calculated in the Chern-Simons theory
and $\nu_L[\Gamma]$ is the {\it linking number} of the path $\Gamma$.
~From here it follows that processes involving the exchange of one pair of
particles
differ by the  phase factor $\exp(i \delta \Delta \nu_L[\Gamma])$, where
we have defined the {\it statistical angle} $\delta={\frac{1}{2\theta}}$
and $\Delta \nu_L[\Gamma]$ is the change of linking number. For an
exchange process, $\Delta \nu_L[\Gamma]=\pm 1$ and the phase factor is
just $\exp(\pm i \delta)$. Thus, $\delta$ is regarded as the change (or
shift) of statistics. By generalizing these arguments for an $N$-particle
system we conclude that
a system of {\it fermions} coupled to a Chern-Simons gauge field with
coupling constant $\theta$ behaves like a system of {\it anyons} with
statistical angle
$\delta= {1\over 2 \theta}$, measured with respect to Fermi statistics
 \cite{e1}. If $\theta={1\over 2 \pi}{1\over 2s}$, where $s$  is an
arbitrary integer, then $\delta=2 \pi s$ and the system still represents
fermions.

In reference \cite{lopez1} we presented a detailed proof of the
physical equivalence of two theories of particles coupled to a
Chern-Simons gauge
field with coupling constants $\theta$ and $\theta \prime$ such that
\begin{equation}
{1\over \theta'}= {1\over \theta} + 2 \pi \times 2s.
\label{eq:c11}
\end{equation}
where $s$ is an arbitrary integer. There, we showed that the physical
observables have exactly the same matrix elements in both theories
and, consequently, that the theories are physically equivalent.
In particular, a
theory of interacting {\it fermions} is always equivalent to a family of
theories of interacting {\it fermions} coupled to a Chern-Simons gauge
field with coupling constant $\theta$ such that ${1\over \theta}=2 \pi \times
2s$. This result is the starting point of our analysis of the FQHE.

~From Eq~\ref{eq:wl}, it is apparent that the physics of these theories
is a {\it periodic} function of the  statistical angle,
$\delta \to \delta + 2 \pi \times {\rm integer}$.
Even though these theories are exactly periodic,
periodicity is broken in approximate, semiclassical approximations, such as
in the Average Field Approximation that we will use here since
{\it one particular period will have to be
chosen}.  This is an important issue since, at the perturbative level,
the fluctuations around this mean-field are not able to restore
periodicity as an exact property. The restoration of periodicity is a
non-perturbative effect. In particular, operators
which create excitations that change the amount of flux per particle in
{\it even} multiples of the flux quantum are soliton operators which restore
the periodicity broken by the mean-field.
However, we can take advantage of the periodicity of the CS
description to choose the period in which the mean-field-theory is
simplest. This is the approach we take in this work to attack the
FQHE. As it will be apparent in the next Section, at the mean-field
level, this approach reproduces Jain's construction of the FQHE
states. Recently, Kivelson, Lee and Zhang \cite{phase} have forcefully argued
that periodicity plays a central role in the determination
of the global phase diagram of the 2DEG in a strong magnetic field.

In second quantized language, what we have proven is the physical
equivalence of
theories whose dynamics are governed by the actions ( in units in which $
\hbar=1$)
${\cal S}_{\theta}$ and ${\cal S}_{\theta '}$ which are defined by
\begin{eqnarray}
{\cal S}_{\theta}&=&\int  d^3 z \; \left\{ \psi^*(z) [i D_0 +\mu]
\psi(z) -{1 \over 2 M}|{\vec D}\psi(z)|^2
+ {\theta \over 4} \epsilon_{\mu \nu \lambda}
a^{\mu}{\cal F}^{\nu \lambda} \right\} \nonumber \\
& -& {1\over 2} \int  d^3z \int d^3z'\;(|\psi(z)|^2-{\bar
\rho}) V(|{\vec z}-{\vec z}'|) (|\psi(z')|^2-{\bar\rho}).
\label{eq:c12}
\end{eqnarray}
where ${\bar \rho}$ is the average particle density, provided that
$\theta$ and $\theta'$ satisfy Eq.~\ref{eq:c11}. In Eq.~\ref{eq:c12}
$\psi(z)$ is a second quantized Fermi field, $\mu$ is the chemical
potential and $D_{\mu}$ is the covariant derivative which couples the fermions
to both the external electromagnetic field $A_{\mu}$ and to the statistical
gauge field $a_{\mu}$
\begin{equation}
D_{\mu}=\partial_{\mu}+i{e\over c}A_{\mu}+ia_{\mu}.
\label{eq:c13}
\end{equation}
In particular, a theory of interacting {\it fermions} ( which has $s=0$)
is equivalent to a {\it family} of theories of {\it fermions} with
${1\over \theta}=2 \pi \times$ even integer.

\subsection{The Semiclassical Limit and The Laughlin Ground State}
\label{subsec:semiclas}

In the remaining of this  Section we will show that the
{\it semiclassical limit} of the theory described by the action
$S_{\theta}$ of Eq.~\ref{eq:c12}, with ${1\over\theta}=2
\pi (m-1)$, yields the same physics as the Laughlin state.
In order to prove this statement we will develop a semiclassical
approach to this problem. As a by-product, our formalism provides for a
systematic procedure to compute corrections to the Laughlin
approximation. This is, to the best of our knowledge, the first formalism for
which the Laughlin {\it ansatz} arises as the first of a series of
approximations.

Consider the quantum partition function for this problem ( at $T=0$)
\begin{equation}
{\cal Z}=\int {\cal D} \psi^* {\cal D} \psi {\cal D} a_{\mu}
\exp (i S_{\theta}).
\label{eq:c14}
\end{equation}
We will treat this path-integral in the semiclassical approximation. In order
to do that, we will first integrate-out the fermions and treat the resulting
bosonic theory within the Saddle Point Expansion (SPE) characteristic of
semiclassical
approaches to quantum mechanics and quantum field theory
\cite{coleman,rajaraman}. In terms of Feynman diagrams, the
semiclassical expansion is an expansion in powers of the number of
fermion loops (bubbles). Thus, the leading semiclassical approximation
(gaussian fluctuations) is equivalent to the conventional
RPA of many-body physics. It is easy to see that the formal
expansion parameter that organizes the SPE is the filling fraction
$\nu$. This is not surprising since RPA is a high density
approximation \cite{bohm-pines}. However, for the FQHE, the filling fraction
$\nu$ is never large but typically  of order one. We will show below
that, in spite of this, the SPE will still be very useful
since, for special values of the filling fraction,
it will enable us to construct a set of ground states for which the
spectrum is fully gapped. Thus, for those states the SPE yields a
perturbative expansion free of infrared divergencies. However we will
also encounter a set of fractions for which there is no gap and the SPE
(or equivalently, RPA) is infrared divergent at every order.
Interestingly enough, the set of fractions for which the SPE predicts a
fully gapped state coincides with the main fractions of the FQHE. These
are the states in Jain's hierarchy.

In the absence of electron-electron interactions, the fermions can
be integrated out immediately since the action becomes quadratic in Fermi
fields.
In the presence of interactions this is no longer possible since the
interaction term in the action Eq.~\ref{eq:c12} spoils this
feature. However, the physical states used in deriving the path integral
{\it must} obey the local charge-flux constraint of the Chern-Simons theory.
Hence, it is legitimate to replace the charge density $j_0(x)$ by
$\theta {\cal B}(x)$ in the pair-interaction term of
the action, at all points of space-time $x$ and
to write the pair-interaction term of Eq.~\ref{eq:c12} in the form
\begin{equation}
{\cal S}_{\rm int}= {1\over 2} \int d^3z \int d^3z'
(\theta {\cal B}(z)-{\bar \rho}) V(z-z') (\theta {\cal B}(z')-{\bar \rho}))
\label{eq:c15}
\end{equation}
where $V(z-z')$ represents the instantaneous pair interaction {\it i.~e.\/},
\begin{equation}
V(z-z')=V(|{\vec z}-{\vec z'}|) \delta(t-t').
\label{eq:c16}
\end{equation}
We will assume that the pair potential has either the Coulomb
form, {\it i.~e.\/}, $V(|{\vec r}|) = {q^2 \over r}$, or that in
momentum space it
satisfies that ${\tilde V}({\vec Q}) {\vec Q}^2$ vanishes at zero momentum.
This includes the case of ultralocal potentials, ({\it i.~e.\/}, with a range
smaller or of the same order as the cyclotron length $\ell$), in which case
we can set ${\tilde V}=0$, or short range potentials with a range longer that
$\ell$ such as a Yukawa interaction.

Note that since we are dealing with a gauge theory, a gauge has to
be specified in order to make the functional integral well defined. We will
assume from now on that a gauge fixing condition has been imposed but, for the
moment, we will not make any specific choice of gauge.

The partition function ${\cal Z}$ can be written in the form of a functional
integral involving
the Fermi fields $\psi$, and the statistical gauge fields $a_{\mu}$.
The Fermi fields can be integrated out without any difficulty yielding, as
usual, a fermion determinant. The resulting partition function can thus be
written in terms of an effective action $S_{\rm eff}$ given by
\begin{eqnarray}
S_{\rm eff}\!\!\!\!\!\!&&=-i {\rm tr} \log [iD_0+ \mu + {1\over 2m}{\vec D}^2]+
 S_{\rm CS}(a_{\mu}-{\tilde A}_{\mu})
\nonumber \\
\!\!\!\!\!\!&&-{1\over 2} \int d^3z \int d^3z'\;
          [\theta ({\cal B}(z)- {\tilde B}(z))-{\bar \rho}]\;
 V(z-z')\;  [\theta ({\cal B}(z')- {\tilde B}(z'))-{\bar \rho}]
\nonumber \\
&&
\label{eq:c21}
\end{eqnarray}
where $D_0$ and ${\vec D}$ are the covariant derivatives of
Eq.~\ref{eq:c13} and
$S_{\rm CS}$ is the Chern-Simons action Eq.~\ref{eq:c5}.
The field ${\tilde A_{\mu}}$
represents a small fluctuating electromagnetic field, with vanishing average
everywhere, which will be used to probe the system. The electromagnetic
currents will be calculated as first derivatives of ${\cal Z}$ with respect to
${\tilde A_{\mu}}$. The full electromagnetic response will be obtained in this
way. Notice that we have used the invariance of the measure
${\cal D} a_{\mu}$ with respect to shifts, to move ${\tilde A}_{\mu}$
out of the covariant derivatives and into the Chern-Simons term $S_{\rm CS}$.

We are now ready to proceed with the semiclassical approximation. The
Saddle Point Equations, or classical equations of motion are
\begin{equation}
{\delta S_{\rm eff} \over \delta a_{\mu}(z)} \vert_{{\bar a}}=0
\label{eq:c22}
\end{equation}
By varying $S_{\rm eff}$ with respect to $a_{\mu}(z)$ we find
\begin{eqnarray}
\langle j_0(z) \rangle_F &=& - {\theta}
 [ \langle {\cal B}(z) \rangle - \langle {\tilde B}(z) \rangle ] \nonumber \\
\langle j_{k}(z)\rangle_F &=& +\theta \epsilon_{kl}
 (\langle {\cal E}_{l}(z) \rangle -  \langle {\tilde E}_{l} \rangle)
 \nonumber\\
&& - \theta \epsilon_{kl}{\partial }^{(z)}_{l}
 \int d^3z' V(|{\vec z}-{\vec z}') \; \big( \theta (\langle {\cal B}-{\tilde B}
 \rangle (z'))-{\bar \rho} \big) \big\}
\label{eq:c23}
\end{eqnarray}
where $\langle j_{\mu}(z)\rangle_F$ represents the expectation value of
the charge and current of the equivalent fermion problem.

These equations have many possible solutions which include uniform
(liquid) states, Wigner crystals, and non-uniform states with vortex-like
configurations. We will only consider solutions with uniform particle
density ($\langle j_0(z) \rangle={\bar \rho}$), {\it i.e.}, the liquid
phase  solution, and no currents in the ground state.

If the external electromagnetic fluctuation is assumed
to have zero average, the only possible solutions of this type are
\begin{eqnarray}
\langle {\cal B} \rangle &=& -{{\bar \rho}\over \theta}\nonumber \\
  \langle {\vec {\cal E}} \rangle &=& 0
\label{eq:c26}
\end{eqnarray}
This is the {\it average field approximation} (AFA) which can be regarded
as a mean field approximation. Eq.~\ref{eq:c26} shows that, for a
translationally invariant ground state, the effect
of the statistical gauge fields, at the  level of the
saddle-point-approximation, is to {\it change} the effective flux
experienced by the fermions. The total effective field is thus reduced from
the value of the external field $B$ down to $B_{\rm eff}=B+\langle {\cal B}
\rangle = B -{{\bar\rho}\over \theta}$. We want to find the ground state
and the spectrum of low energy excitations of a system of
$N$ (interacting) electrons in the presence
of an external magnetic field of strength $B$. We will further assume that the
linear size $L$ of the sample is such that a total of $N_{\phi}$ quanta of the
magnetic flux are piercing the surface. In general, the filling fraction
$\nu={N\over N_{\phi}}$ is not an integer. Notice that, as $B$ and
$\rho$ are varied, the effective field $B_{\rm eff}$ may either be
parallel {\it or} anti-parallel to the external field $B$. Thus, we will
not assume a particular sign for $B_{\rm eff}$ although we will set
$B>0$.

The uniform effective magnetic field $B_{\rm eff}$, which solves
Eq.~\ref{eq:c26}, define a new set of {\it effective} Landau levels.
Each level has a degeneracy equal to the total number of {\it
effective } flux quanta
$N_{\rm eff}$ and the separation between levels is the effective
cyclotron frequency $\omega_c^{\rm eff}={e |B_{\rm eff}|\over M c}$.
Similarly, there is an effective cyclotron radius $\ell^{\rm eff}$.
It is easy to see that the uniform saddle-point state which satisfies
Eq.~\ref{eq:c26} has a gap {\it only if} the effective
field $B_{\rm eff}$ experienced by the $N$ fermions is such that the
{\it fermions fill exactly an integer number $p$ of the effective Landau
levels}. This is precisely the point of view advocated by Jain: the FQHE is an
IQHE of a system of electrons dressed by an even number of flux quanta.
However, this condition cannot be met for arbitrary values of the filling
fraction $\nu$ at fixed field ( or at fixed density). Let $N_{\phi}^{\rm eff}$
denote
the effective number of flux quanta piercing the surface after
screening. It is given by
\begin{equation}
\pm 2 \pi N_{\phi}^{\rm eff}=2 \pi N_{\phi}-{{\bar \rho}\over
\theta}L^2.
\label{eq:c27}
\end{equation}
where the $\pm$ sign stands for the case of an effective field parallel
or antiparallel to $B$.
Thus, the effective cyclotron frequency $\omega_c^{\rm eff}$ is {\it
reduced} from its free electron value of ${e B\over M c}$ down to
$\omega_c^{\rm eff}=\omega_c(1-{\nu \over 2 \pi \theta})$. The effective
cyclotron radius is given by $\ell^{\rm eff}=(\ell/
{\sqrt{1-{\nu \over 2 \pi \theta}}})$ which is larger than the
non-interacting value. Therefore, even though the bare Landau levels may
be separated by a sizable Landau gap $\hbar \omega_c$, the effective Landau
levels have the smaller gap $\hbar \omega_c^{\rm eff}$.

Substituting the value of $\theta$ in  Eq.~\ref{eq:c27} we obtain
\begin{equation}
\pm 2 \pi N_{\phi}^{\rm eff}= 2 \pi N_{\phi}-2 \pi 2s N.
\label{eq:c28}
\end{equation}
where $2s$ is an even integer.
The spectrum supported by this state has an energy gap if the $N$ fermions
fill exactly $p$
of the Landau levels created by the effective field $B_{\rm eff}$. In other
words, the {\it effective} filling fraction is $\nu_{\rm eff} \equiv {N\over
N_{\phi}^{\rm eff}} =p$. Using Eq.~\ref{eq:c28}, we find that the
filling fraction $\nu$ and
the external magnetic field $B$ must satisfy
\begin{equation}
\pm {N\over p}={N\over \nu}-2s N,
\label{eq:c29}
\end{equation}
or, equivalently,the allowed filling fractions are
\begin{equation}
\nu_\pm(p,s)={\frac{p}{2sp \pm 1}}
\label{eq:c30}
\end{equation}
The allowed filling {\it fractions} $\nu_\pm$ can thus be arranged into
families or hierarchies. As $p \to \infty$, the allowed fractions approach
the limiting values $\lim_{p\to \infty} \nu_\pm(p,s)={\frac{ 1}{2s}}$,
with $\nu_+(p,s)\leq {\frac {1}{2s}}$ while $\nu_-(p,s)\geq {\frac {1}{2s}}$.
For the special case $s=1$, the limiting value is
${\frac{1}{2}}$. Since $\nu_+(p,1)\leq {\frac {1}{2}}$ and
$\nu_-(p,1)\geq {\frac {1}{2}}$, the sequence $\nu_+(p,1)$ is said to
represent {\it electron}-like FQH states whereas the mirror sequence
$\nu_-(p,1)$ is said to represent {\it hole}-like FQH states.
The effective Landau gap for these solutions is
\begin{equation}
\hbar \omega_c^{\rm eff}={\hbar \omega_c\over 2sp\pm 1}
\label{eq:c32}
\end{equation}
which is small if either $p$ or $s$ are large. Thus, the energy to
excite a {\it fermion} is $\hbar \omega_c^{\rm eff}$ and it is considerably
smaller than the bare free-particle value $\hbar \omega_c$.

The states are thus parametrized by two integers $p$ ( the number of filled
{\it effective} Landau levels of the effective field) and $2s$
( the number of flux quanta
carried by each fermion), and by a sign. It is trivial to see that
the Laughlin sequence is an obvious solution of
Eq.~\ref{eq:c30}  for $p=1$ and $2s=m-1$. The effective fermions thus
fill up exactly one Landau level and $\theta$ has to be chosen to be
${1\over \theta}=2 \pi (m-1)$. This is  Jain's result. At this
mean-filled level the wave function is the Slater determinant for one
filled Landau level $\chi_1$ of Eq.~\ref{eq:c3}. We will
show below that the additional factor $\prod_{i<j}(z_i-z_j)^{m-1}$ is due to
fluctuations. For general $p$ and $s$, these are the allowed states in
Jain's hierarchy \cite{jain1,jain2}. Please notice that, unless
$s=0$, all states in these hierarchies have $\nu_\pm(p,s) \leq 1$. For
$s=0$ ({\it i.~e.\/}, no flux attachement) $\nu=p$ and we reproduce the
integer QH states. Thus, only {\it primary} FQH states are generated.
States characterized by other fractions can be constructed by means of
the conventional scheme of {\it condensing} quasiparticles or quasiholes
on top of these sequencies. We will see below that, as expected, the
quasiparticles and quasiholes of the FQH states in these hierarchies are
{\it anyons}. However, it is unclear how many of these higher level
hierarchical states are energetically stable (or accessible).

The saddle-point approximation yields a very simple spectrum which
consists just in the single and many-particle excitations of fermions in
the effective field $B_{\rm eff}$. The single-particle gap is
equal to $\hbar \omega_c^{\rm eff}=\hbar \omega_c/(2sp\pm 1)$.
Thus, at fixed $B$, the single particle energy gaps become smaller as the
index $p$ increases and vanish as $p \to \infty$.
In this limit the hierarchical states converge to
the fractions $\nu={\frac{1}{2s}}$. For these fractions
the system is compressible, {\it i.~e.\/}, there is no energy gap in its
spectrum,
and the perturbation expansion breaks down. The
breakdown is signalled by infrared divergent contributions at every order
of the expansion at low temperatures.
The case of the ${\frac{1}{2s}}$ states was analyzed by
Halperin, Lee and Read \cite{hlr}. Elsewhere in this volume, Steve Simon
presents a detailed discussion of this important case.

At this level of the SPE, the Coulomb energy does not enter explicitly in the
effective single particle gaps. However, their existence is a physical
consequence of the existence of Coulomb interactions. Moreover,
in the strong field limit $B \to \infty$, both the bare and the effective
Landau gaps diverge. This is clearly incorrect since, as $B$ increases
the states of the system should be made almost completely of states
only in the lowest Landau level. In the strong field limit, the
Coulomb energy alone should determine the numerical value of the gaps.
The origin of this difficulty is that the SPE (or the RPA) is formally
correct in the small field limit and the states exhibiting the FQHE are all
in the large field limit. Thus, we should expect that the higher order
correction in the SPE should not only correct the numerical value of the
gaps but also lead to a finite limit in the strong field regime. These
corrections have not been calculated and their properties are not yet
understood. The same considerations apply to the spectrum of collective
excitations discussed in Section \ref{subsec:resfun}. However, it is
important to stress that the SPE (with gaussian corrections) yields  a
set of states with the correct quantum numbers even though the
energies are not correct. In particular we will show below that, once
gaussian fluctuation are accounted for, these single particle states
have the correct charge and statistics (although the wrong energy).

The saddle-point equations, Eq.~\ref{eq:c23}, have a host of {\it
non-uniform} solutions which have finite energy.
They can be viewed as {\it
soliton} or {\it vortex} solutions. Our construction is very close in
spirit to the picture of the quasihole presented by
Laughlin \cite{laughlin1}.
We will only discuss here the qualitative features of these solutions.
For the sake of simplicity, we will only study the case of fermions in the
Laughlin $1\over m$ sequence.
Recall that this sequence is represented by the solution with $p=1$ and
$m=2s+1$. The uniform state was constructed by filling up the lowest
effective Landau level. Let us consider the state which results from
removing a fermion from the single-particle state, centered around the
origin $z=0$ and with lowest angular
momentum, and placing it on the first unoccupied angular momentum state.
Physically, this new state lies on the outer edge of the system.
For an uniform effective field, this state does not exist. But, if the
effective field is {\it increased} at the origin by an amount equal to
one flux quantum, the angular momentum of all its eigenstates is
raised by one whole unit. The radius $R_N$ of the droplet with $N$
particles swells to a new value $R_N+\delta R$ large enough to include
a new cyclotron orbit. Qualitatively, a quasihole localized at $z_0$ has
the {\it mean-field} wave function $\Psi_h=\Psi_h(z_0,\{ z_j \})$ given
by
\begin{equation}
\Psi_h(z_0,\{ z_j \})=\prod_{i=1}^N(z_i-z_0) \; \chi_1( \{ z_j \}).
\label{eq:c36}
\end{equation}
where $\chi_1( \{ z_j \} )$ is the wave function for $N$ fermions
occupying the lowest Landau level. Notice that this wave function
differs from the Laughlin state for the quasihole by the prefactor
$\prod_{i<j}(z_i-z_j)^{m-1}$. Indeed, this prefactor is also missing in
the {\it mean-field} wave function for the ground state. In our picture,
both prefactors arise from fluctuations which attach fluxes to the
particles. The quasihole wave function of Eq.~\ref{eq:c36} is an
approximation valid in the limit $|z_i-z_0| \gg \ell^{\rm eff}$. From
Eq.~\ref{eq:c23} it is easy to see that the mean-field
{\it excitation energy} of the quasihole
$\varepsilon_h$ is given approximately by the Coulomb energy
$V(\ell^{\rm eff})$.

\subsection{The role of the Gaussian fluctuations}
\label{subsec:gaussian}

We consider now the role of the gaussian fluctuations around
the {\it uniform} classical solutions discussed in Section
\ref{subsec:semiclas}. We begin by considering
the effective action of Eq.~\ref{eq:c21}.
Let ${\tilde a}_{\mu}(x)$ denote the {\it fluctuations} of the
statistical vector potential $a_{\mu}$, {\it i.~e.\/}, we set
$a_{\mu} \to \langle a_{\mu}\rangle + {\tilde a}_{\mu}$.
The effective action of Eq.~\ref{eq:c21} can be expanded in a series
in powers of the fluctuations. We will be interested only in keeping
just up to quadratic terms in the fluctuations. In the language of
Feynman diagrams, we are summing up all the one-loop bubble
contributions. Thus, the effective action at the quadratic level
involves the linear response kernels (evaluated in the RPA) for a
system of fermions in an external static uniform magnetic field
$B_{\rm eff}$ with an integer number of filled effective Landau
levels.  As usual, the linear terms are cancelled if the saddle-point
equations are satisfied.

At the quadratic (gaussian) level the effective action has the form
\begin{equation}
S^{(2)}={1\over 2} \int d^3x d^3y\; {\tilde  a}_{\mu}(x)\;
\Pi^{\mu \nu}(x,y)\; {\tilde a}_{\nu}(y)
 + S_{\rm CS}({\tilde {a}}_{\mu}-{\tilde  A}_{\mu})+
S_{\rm int}^{(2)}({\tilde { a}}_{\mu}-{\tilde  A}_{\mu}).
\label{eq:c40}
\end{equation}
where
\begin{equation}
S_{\rm int }^{(2)}({\tilde a}_{\mu}-{\tilde  A}_{\mu})=
-{ \theta^2 \over 2} \int d^3 x d^3 y
\left[ {\tilde {\cal B}}(x) - {\tilde B}(x) \right]  V({\vec x}-{\vec y})
 \left[ {\tilde {\cal B}}(y) - {\tilde B}(y) \right]
\label{eq:c38}
\end{equation}
Notice that the only approximation used in Eq.~\ref{eq:c40} is in the first
term which follows from the expansion of the fermion determinant in
powers of ${\tilde a}_{\mu}$. The remaining terms are exact.

Eq.~\ref{eq:c40} holds provided that the non-quadratic dependence in the
fluctuating part of the statistical vector potentials ${\tilde a}_\mu$
is small. Recall that these non-quadratic terms result from expanding the
(logarithm) of the fermion determinant in powers of the fluctuations
around the average field approximation. The kernels that enter in the
expressions for these terms are (connected) current correlation
functions (or response functions) of the mean field theory. Thus,
the tensor $\Pi^{(0)}_{\mu\nu}$ is the {\it polarization tensor} of the
equivalent fermion problem at the mean field level, and it is obtained
by expanding the fermion determinant up to quadratic order in the statistical
gauge field
\begin{eqnarray}
 {\cal Z}_0[A_{\mu}]&\equiv& {\rm Det}\; [iD_0+ \mu + \lambda +
{1\over 2M}{\vec D}^2]\nonumber \\
 &=& {\cal Z}[0] \; \exp \{{i\over 2}
                  \int d^{D}x \int d^{D}y \; A_{\mu}(x) \Pi^{(0)}_{\mu \nu}
                  (x,y) A_{\nu} (y)+ \ldots  \}
\label{eq:c41}
\end{eqnarray}
The tensor $\Pi^{(0)}_{\mu\nu}$ {\it should not}
be confused with the true {\it electromagnetic} polarization tensor
which measures the response of the whole system to a weak
electromagnetic field. We will come back to this issue in the next Section.

It is straightforward to derive an expression for $\Pi^{(0)}_{\mu\nu}$ in terms
of the one-particle Green functions $G(x,y)$
\begin{equation}
G(x,y) =<x|{1 \over i D_0 + {\mu} + {1\over 2M} {\vec D}^2[<A>]}|y>.
\label{eq:c42}
\end{equation}
The components of the polarization tensor $\Pi^{(0)}_{\mu \nu}(x,y)$
are  $(\hbar = 1)$
\begin{eqnarray}
\Pi ^{(0)}_{00} (x,y)=\!\!\!\!\!\!\!\! && i \ G(x,y) G(y,x) \nonumber \\
\Pi^{(0)} _{0j} (x,y)=\!\!\!\!\!\!\!\!&& {1 \over 2M}
                    \{ G(x,y) D_j ^y G(y,x) - G(y,x)
                    D_j ^{y \dagger } G(x,y) \} \nonumber \\
\Pi^{(0)} _{j0} (x,y)=\!\!\!\!\!\!\!\!\!&& +{1 \over 2M} \{ -G(x,y)
                    D_j ^{x \dagger} G(y,x)
                    + G(y,x) D_j ^x G(x,y) \} \nonumber \\
\Pi^{(0)} _{jk} (x,y)=\!\!\!\!\!\!\!\!\! &&{i \over M} \delta (x-y)
                    \delta _{jk} G(x,y) +      \nonumber \\
        -&&\!\!\!\!\!\!\!\!\!  {i \over 4M^2} (D_j ^x G(x,y)) (D_k ^y G(y,x))
                -{i \over 4M^2} (D_j ^{x \dagger } G(y,x))
                    (D_k ^{y \dagger } G(x,y)) + \nonumber \\
          +\!\!\!\!\!\!\!\!&&{i \over 4M^2} G(y,x)) (D_j ^x D_k ^{y \dagger }
                    G(x,y))
                  +{i \over 4M^2} (D_j ^{x \dagger } D_k ^y
                    G(y,x)) G(x,y)  \cdot
\nonumber \\
&&
\label{eq:c43}
\end{eqnarray}
It can be shown that the effective action of Eq.~\ref{eq:c41} is
gauge-invariant and that, in consequence, $\Pi^{(0)}_{\mu\nu}$ is {\it
transverse}, {\it i.~e.\/},
\begin{equation}
\partial_{\mu}^x\Pi^{(0)}_{\mu\nu}(x,y)=0 .
\label{eq:c44}
\end{equation}
In reference \cite{lopez1} we give detailed expressions for the various
components of $\Pi^{(0)}_{\mu\nu}$. In particular we show that the action is
gauge-invariant and that $\Pi^{(0)}_{\mu\nu}$ is transverse, albeit only in
a weak sense {\it i.~e.\/},  as a distribution. Eq.~(\ref{eq:c43})
agrees with the results
of Randjbar-Daemi, Salam and Strathdee \cite{salam} but disagrees
with the calculation of $\Pi^{(0)}_{\mu\nu}$ by Chen {\it et.~ al.\/}
\cite{cwwh}.

~From the expressions for the components of $\Pi^{(0)}_{\mu\nu}$ in
reference \cite{lopez1}, we see that this tensor is analytic in
${\vec Q}^2\over {B_{\rm eff}}$ and that it has simple
poles at  $\omega=k {\bar \omega_c}$ (with $k$ an integer),
where
${\bar \omega_c}\equiv \omega_c/(2sp+1)$ is the cyclotron frequency
associated with the effective magnetic field $B_{\rm eff}$. As a
result, $\Pi^{(0)}_{\mu\nu}$ has a gradient expansion in powers of the inverse
of the effective magnetic field $1\over {B_{\rm eff}}$, or equivalently, in
powers of the inverse of the external magnetic field $1\over B$. In fact, the
dimensionless parameter of this expansion is ${\vec Q}^2\over B$ (we are
working in a system of units such that $\hbar = c =e=1$). It also turns out
that, within this approximation, the limits of $B \rightarrow \infty$ and
$M \rightarrow 0$ are not equivalent (see the explicit form of
$\Pi^{(0)}_{\mu\nu}$ given in reference \cite{lopez1}).

The non-quadratic terms in ${\tilde a}_\mu$ in the effective action are of
the form
\begin{equation}
 S_{\rm eff} =S^{(2)}({\tilde a}^{\mu},{\tilde A}^{\mu})
 + \frac{1}{3!}\int d^3x\; d^3y\; d^3z\;{\tilde a}^{\mu}\;
{\tilde a}^{\nu}\; {\tilde a}^{\lambda}\;
\Pi^{(0)}_{\mu\nu\lambda}(x,y,z)  +\ldots
\end{equation}
where the kernel $\Pi^{(0)}_{\mu\nu\lambda}(x,y,z)$ represents a
three-point current correlation function in the mean field theory. Thus,
in the language of Feynman diagrams, while $\Pi^{(0)}_{\mu\nu}(x,y)$
can be viewed as a fermion bubble with two amputated external collective
mode lines, $\Pi^{(0)}_{\mu\nu\lambda}(x,y,z)$ again has one
fermion loop tied to three amputated external collective mode lines
${\tilde a}_\mu$. Each one of these non-quadratic kernels have the same
gauge invariance ({\it {\it i.~e.\/}}, transversality) and analytic properties
as the
gaussian (or RPA) kernel. In particular, this means that, in momentum
and frequency space, these kernels must be a linear combination of
tensors (of the appropriate rank) which have the correct transversality
properties, times a set of functions which are analytic in ${\vec Q}^2$ and
have
poles at frequencies equal to an integer multiple of the effective
cyclotron frequency. Therefore, the non-quadratic terms necessarily have powers
of ${\vec Q}^2 \over {B_{\rm eff}}$ (for each one of the external momenta and
frequency entering the fermion loop) which are higher than the ones
found at the quadratic level. Since the mean field theory has an integer
number of filled Landau levels, the energy denominators of the kernels do not
change this counting in powers of ${\vec Q}^2 \over {B_{\rm eff}}$.
In conclusion, the expansion of the fermion determinant, and hence of the
effective action, is actually an expansion in powers of
${\vec Q}^2\over B_{\rm eff}$, or equivalently, in powers of
${\vec Q}^2\over B$. However, an expansion in
powers of ${\vec Q}^2\over B$ is also a gradient expansion. Thus, the
gradient expansion and the semiclassical expansion mix and are not
independent from each other.

The semiclassical expansion is obtained according to the following
rules. The propagator for the fluctuations, which represent collective
modes, is the inverse of the kernel of the gaussian action.
Since the pair
potential enters only through the propagator for the fluctuations,
the perturbation theory is not
an expansion in the powers of the pair interaction.
From
this point of view, this expansion is very different from conventional
expansions around the Hartree  and Hartree-Fock approximations. The
vertices of the expansion are the kernels for the non-quadratic terms.
This expansion lacks a natural small parameter ({\it {\it i.~e.\/}}, a coupling
constant) and it should be regarded, like all semiclassical expansions,
as an expansion in the number
of fermion loops ({\it {\it i.~e.\/}}, RPA plus corrections). One should keep
in mind,
however, our previous discussion on its exactness in powers of
${{\vec Q}^2 \over B}$. In what follows
we will make extensive use of the formal properties of this expansion.

We remark here that, unlike other mean field approaches to
conventional many body systems (such as Hartree-Fock), the gaussian corrections
of this theory {\it must} alter the
{\it qualitative} properties of the state described by the AFA. The
reason is that the AFA violates
explicitly space-time symmetries, such as Galilean invariance (more
generally, {\it magnetic invariance}) which, for translationally
invariant systems, must remain unbroken and unchanged. Thus the center
of mass of the system must execute a cyclotron-like motion at, exactly, the
cyclotron frequency of non interacting electrons in the full external magnetic
field, as demanded by Kohn's theorem \cite{kohn}. A na{\"\i}ve
application of the AFA would suggest that the cyclotron frequency is
renormalized downwards since the effective field seen by the composite
fermions is smaller than the external field $B$. Hence, the {\it
magnetic algebra} may appear to have changed. We will see below that the
gaussian fluctuations yield the correct cyclotron frequency and,  thus,
restore the correct magnetic algebra.

\section{Response functions and spectrum of collective excitations}
\label{sec:gaflu}

The physical systems which exhibit the FQHE present a very rich response
to external electromagnetic
perturbations. While some of the observed phenomena, such as cyclotron
resonance, can be understood in terms of simple global motions of the
center of mass under the combined influence of electric and magnetic
fields, the spectrum of collective excitations is certainly determined
by the interactions. Given the unusual features of the Laughlin states
and its generalizations, it is expected that the same features should
largely determine the behavior of the collective modes too.
We will now use the fermion field theory described in
Section \ref{sec:csfqhe}, to
study the collective excitations of the fully polarized FQHE states in
the sequence $1/\nu=\pm 1/p+2s$ (where $p$ and $s$ are two positive
integers).
While the average field  approximation is certainly
very appealing, it has the serious problem that it breaks a number of
space-time symmetries quite explicitly. In particular, it breaks both
Galilean and Magnetic invariance. It turns out that the leading quantum
fluctuations around this state, restore these symmetries, in the
uniform ${\vec Q} \to 0$ limit, already at the gaussian level. Indeed,
we find that the quadratic or gaussian level of the semiclassical
expansion gives the correct value of the Hall conductance of the
system, and that the leading
order of the density correlation function saturates the {\it f}-sum rule.
This is an essential result to show that the absolute value squared of the
wave function of all the (incompressible) liquid states has the Laughlin
form at very long distances, in the thermodynamic limit (see
Section \ref{sec:wave}).

\subsection{Electromagnetic response functions for the FQHE}
\label{subsec:resfun}

The effective action for the
fluctuations of the Chern-Simons gauge field (${\tilde a}_{\mu}$), at the
Gaussian level of our approximation, is given by
\begin{eqnarray}
S^{(2)}({\tilde a}^{\mu},{\tilde A}^{\mu})
             &=&{1\over 2} \int d^3x \; d^3y\; {\tilde a}^{\mu}(x)\;
             \Pi^{(0)}_{\mu\nu}(x,y) \; {\tilde a}^{\nu}(y) +
       {\theta \over 4} S_{\rm cs}({\tilde a}_{\mu}-{\tilde A}_{\mu})
\nonumber \\
&-&{\theta^2 \over 2} \int d^3 x \; d^3 y \;  \left[ {\tilde {\cal B}}(x)-
{\tilde B} (x) \right] \; V(x-y) \;
 \left[ {\tilde {\cal B}}(y)- {\tilde B} (y) \right]
 \nonumber \\
&&
\label{eq:r9}
\end{eqnarray}
Since $S^{(2)}$ is quadratic in ${\tilde a}_{\mu}$, we can integrate out this
field and obtain the effective action for the electromagnetic fluctuations
${\tilde A}_{\mu}$, $S_{\rm eff}^{\rm em}({\tilde A}_{\mu})$ . We will
use this effective action to calculate the full electromagnetic response
functions at the gaussian level. Since this calculation is based on a
one loop effective action for the fermions ({\it {\it i.~e.\/}},
a sum of fermion bubble diagrams), this approximation amounts
to a random phase correction to the average field approximation.

In order to integrate out the statistical gauge
field ${\tilde a}_{\mu}$ we must fix the gauge. The
electromagnetic effective action, being gauge invariant, is
independent of the choice of gauge for the statistical gauge fields in
the path integral. We fix the gauge $\partial _{\mu} {\tilde a}^{\mu}={\alpha}$
 (where $\alpha$ is an arbitrary real number)  using the standard
Faddeev-Popov procedure
\footnote{To be more precise, whenever we have had to
fix the gauge we have adopted Feynman's method of averaging
over gauges, with $\alpha$ being the width of the
distributions. This is a standard procedure which is reviewed in many
textbooks (see, for instance, reference \cite{ramond}).}.
The result is explicitly gauge invariant and
all dependence on the parameter $\alpha$ cancels out. At the one-loop
level (governed by the effective action of Eq.~\ref{eq:r9})
we need to know the inverse of the polarization tensor
of the equivalent fermion problem, $\Pi^{(0)}_{\mu\nu}$. In
reference \cite{lopez1}, we showed
that $\Pi^{(0)}_{\mu\nu}$ can be written in terms of three gauge invariant
tensors, an ${\vec E}^2$ term, a ${\vec B}^2$ term, and a Chern-Simons term.
These three tensors plus  $ B {\vec{\nabla}}.{\vec E}$ and a gauge
fixing term (such as ${1\over 2\alpha}({\partial _{\mu}}{{\tilde a}^{\mu}})^2$
which corresponds to the Landau-Lorentz gauge if $\alpha \rightarrow 0$),
close an algebra that
can be used to invert the polarization tensor and to calculate explicitly the
electromagnetic response functions.

After integrating out the statistical gauge field in Eq.~\ref{eq:r9},
the effective action for the electromagnetic fluctuations
${\tilde A}_{\mu}$ turns out to be
\begin{equation}
{\cal S}_{\rm eff}^{\rm em} ({\tilde A}_{\mu}) = { 1\over 2}
                  \int d^{3}x d^{3}y {\tilde A}_{\mu}(x) K^{\mu \nu}(x,y)
                   {\tilde A}_{\nu} (y)
\label{eq:r11}
\end{equation}
Here $K^{\mu\nu}$ is the electromagnetic polarization tensor. It measures the
linear response of the system to a weak electromagnetic perturbation.
Its components can be written in momentum space as follows
\begin{eqnarray}
 K_{00} &=& {\vec Q}^2  K_{0}(\omega, {\vec Q}) \nonumber \\
 K_{0j} &=& {\omega} Q_{j} K_{0}(\omega, {\vec Q})
             + i {\epsilon _{jk}}Q_{k} K_{1}(\omega, {\vec Q}) \nonumber \\
 K_{j0} &=& {\omega} Q_{j} K_{0}(\omega, {\vec Q})
             - i {\epsilon _{jk}} Q_{k} K_{1}(\omega, {\vec Q}) \nonumber \\
 K_{ij} &=& {\omega}^2 {\delta _{ij}} K_{0}(\omega, {\vec Q})
             - i {\epsilon _{ij}} {\omega} K_{1}(\omega, {\vec Q})
 + ({\vec Q}^2 {\delta _{ij}}- {Q_i}{Q_j})K_{2}(\omega, {\vec Q})\nonumber \\
&&
\label{eq:r12}
\end{eqnarray}
where the functions $K_{l}(\omega, {\vec Q})$ ($l=0,1,2$) are given by
\begin{eqnarray}
K_{0}(\omega, {\vec Q}) &=& - {\theta ^2} \;
              { \Pi _{0} \over D(\omega,{\vec Q})}
\nonumber\\
K_{1}(\omega, {\vec Q}) &=& \theta
             +{\theta}^2 \; {(\theta +{\Pi _1})
                             \over D(\omega,{\vec Q})}
    +{\theta}^3 {\tilde V}(\vec Q) {\vec Q}^2 \; { {\Pi_{0}}
                             \over  D(\omega,{\vec Q})}
\nonumber\\
K_{2}(\omega, {\vec Q}) &=& -{\theta}^2 \;{{\Pi_2}\over D(\omega,{\vec Q})}
    + {{\tilde V}(\vec Q)\; ({\omega}^2 \;
             {\Pi_0}^2 -{\Pi_1}^2 )\over D(\omega,{\vec Q})}
    + {{\tilde V}(\vec Q)\; {\vec Q}^2\; {\Pi_0} \;
    {\Pi_2} \over D(\omega,{\vec Q})}
\nonumber\\
&&
\label{eq:r13}
\end{eqnarray}
and
\begin{equation}
D(\omega,{\vec Q}) = {\Pi_0}^2{\omega}^2-({\Pi_1}+\theta)^2
    +{\Pi_0}  ({\Pi_2}-{\theta}^2{\tilde V}(\vec Q)){\vec Q}^2
\label{eq:r14}
\end{equation}
The coefficients $\Pi_{l}$ ($l=0,1,2$) are functions of $\omega$ and
${\vec Q}$, and are given explicitly in reference \cite{lopez1}.
${\tilde V}(\vec Q)$ is the Fourier transform of the interparticle pair
potential.
As we mentioned before, we needed to include a gauge fixing term to be able to
compute the functional integral in Eq.~\ref{eq:r9}. But at the end of the
calculation
all the terms which contain the gauge fixing coefficient ($\alpha$) cancel
each other, and the final result for the response functions is, as it must be,
gauge invariant.
The other tensor that we have introduced to make the calculations,
$B {\vec{\nabla}}.{\vec E}$, is not present in the final answer either.

We want to stress here that the thermodynamic limit is crucial for the
accuracy of our results. Notice first that in the electromagnetic
effective action of Eq.~\ref{eq:r11} we are neglecting higher order response
functions, {\it {\it i.~e.\/}}, correlation functions of three or more
currents or densities.
We have shown in Section \ref{subsec:gaussian} that these higher order
correlation functions
have higher order powers of ${{\vec Q}^2 \over B}$ than the quadratic
term. Strictly speaking, these terms are not neglectible for a finite system
because, in this case, there is a minimum value that the momentum can
take,
determined by the linear size of the system $L$, {\it {\it i.~e.\/}},
$|\vec Q| > {1\over L}$. But in the thermodynamic limit,
$L \rightarrow \infty$ and the momentum can go to zero. In other words,
only for an infinite system one is allowed to keep only the quadratic term in
the electromagnetic action, Eq.~\ref{eq:r11}, and to neglect the higher order
correlation functions.

The electromagnetic response functions determined by $K_{\mu\nu}$ have
the following properties:

i) The polarization tensor
at mean field level, $\Pi^{(0)}_{\mu\nu}$, has poles at every value of the
effective cyclotron
frequency (${\bar \omega}_{c}\equiv {B_{\rm eff} \over M}$). This corresponds
to the physical picture, at mean field level, of an IQHE of the bound states
in the presence of a partially screened external magnetic field,
$B_{\rm {eff}}$. Once we take into account the Gaussian fluctuations,
it is easy to prove that all this poles that are present in the numerator and
the denominator of the $K_{\mu\nu}$ components through the $\Pi_{j}$'s, cancel
out, and the poles of the response functions are determined only by the zeroes
of their denominator, $D(\omega, \vec Q)$. In other words, the collective
excitations of this system will be determined only by the zeroes of
$D(\omega, \vec Q)$.

ii) The leading order term in ${\vec Q}^2$ of the $K_{00}$
component of the polarization tensor saturates the $f$-sum rule.

iii) The gaussian fluctuations of the statistical gauge field are
responsible
for the FQHE. In particular, the gaussian corrections yield the exact
value for the Hall conductance.

In the remaining of this section we will discuss these properties and
their experimentally accessible consequences in detail.

\subsection{ The Spectrum of Collective Excitations}
\label{subsec:colectivo}

For simplicity, we give here the zeroes of $D(\omega, \vec Q)$
for the case of $p=1$ filled effective Landau levels.
The (more tedious) case of general $p$ can be studied by straightforward
application of the same methods \cite{lopez3}.

In the case $p=1$ the filling fraction is $\nu = {1\over m}$, where $m=1+2s$,
{\it {\it i.~e.\/}}, the Laughlin sequence.
We find that there is a family of collective modes whose zero-momentum
gap is $k {\bar \omega}_{c}$, where $k$ is an integer number different from
$1$ and $m$, and whose dispersion curve $\omega_{k} ({\vec Q})$ is
\begin{equation}
\omega_{k} ({\vec Q})= \Big [ (k {\bar \omega}_{c})^2
          + ({{\vec Q}^2\over 2 {B_{\rm eff}}})^{k-1}\;
                         {\bar \omega}_{c}^2
 { {2k (m-1)(k-1)} \over {(k-1)!\; (k-m)}} \Big ] ^{1\over2}
\label{eq:r15}
\end{equation}
The residue in $K_{00}$ corresponding to this pole is
\begin{equation}
 Res(K_{00},\omega _{k}({\vec Q})) = -{\vec Q}^2 \;
                                   { \omega}_{c}{\nu \over 2\pi}\;
        ({{\vec Q}^2\over 2 {B_{\rm eff}}})^{k-1}
   \times { {2k (m-1)(k-1)}\over {(k-1)!(k-m)(k^{2}-m^{2})}}
\label{eq:r16}
\end{equation}
The cases $k=1,m$ have to be treated separately. In general, we find
that there is no mode with a  zero momentum gap at ${\bar \omega}_{c}$.
Instead, at ${\vec Q}=0$, there is a doubly degenerate mode with a gap
at
$ \omega_{c}$. This degenerate cyclotron mode can be viewed as the
mixing of the modes with $k=1$ and with $k=m$.
Thus, the mode
with $k=1$ has been ``pushed up" to the cyclotron frequency (at ${\vec
Q}=0$).
For ${\vec Q} \not=0$, the degeneracy is lifted and these two modes have
different dispersion curves.

For the special case of $\nu = {1\over 3}$, {\it {\it i.~e.\/}},
$m=3$, this effect is particularly important. The dispersion relations
for the cyclotron modes are given by
\begin{equation}
\omega_{\pm} ({\vec Q})  = \Big [ {\omega_{c}}^2 +
        ({{\vec Q}^2\over 2 {B_{\rm eff}}}){{\bar \omega}_{c}^2 \over 2}
            \alpha _{\pm }\Big ] ^{1\over 2}
\label{eq:r17}
\end{equation}
where
\begin{equation}
\alpha _{\pm} = 8+ {2M {\tilde V}({0}) \over 2\pi} \pm
 {\Big (} (8+ {2M {\tilde V}(0) \over 2\pi})^2 + 288 {\Big )}^{1\over 2}
\label{eq:r18}
\end{equation}
The residues corresponding to these poles are
\begin{equation}
Res(K_{00},\omega _{\pm }({\vec Q})) = -{\vec Q}^2 \;
                                { \omega}_{c}{\nu \over 2\pi}\;
                                ({1+ {288\over \alpha_{\pm}^2}})^{-1}
\label{eq:r19}
\end{equation}
For  $\nu = {1\over m}$, $m \geq 5$, the collective modes whose
zero-momentum gap is the cyclotron frequency, $\omega _c$, are
\begin{equation}
 {\omega_{+}}(\vec Q)= \Big [{\omega _{c}}^2 +
         ({{\vec Q}^2\over 2 {B_{\rm eff}}}) \; {\bar \omega}_{c}^2
         \times {\Big(} 4 {{(m-1)}\over (m-2)}+
        {2M {\tilde V}(0) \over 2\pi}{\Big)}\Big ] ^{1\over2}
\label{eq:r20}
\end{equation}
with residue
\begin{equation}
Res(K_{00},\omega _{+}({\vec Q})) = -{\vec Q}^2
                                {\omega}_{c}{\nu \over 2\pi}
\label{eq:r21}
\end{equation}
The other cyclotron mode has the dispersion
\begin{equation}
{\omega_{-}}(\vec Q) = \Big[ {\omega _{c}}^2 -
                        ({{\vec Q}^2\over 2 {B_{\rm eff}}})^{m-2}\;
                        {\bar \omega}_{c}^2
   {{4 m^2 (m-1)^2} \over {(m-1)!}}\;  \Big( 4 {{(m-1)}\over (m-2)}+
  {2M {\tilde V}(0) \over 2\pi}\Big)^{-1}      \Big] ^{1\over2}
\label{eq:r22}
\end{equation}
with residue
\begin{eqnarray}
 Res(K_{00},&&\!\!\!\!\!\!\!\!\!\omega _{-}({\vec Q}))  =\nonumber\\
&&\!\!\!\!\!\!\!\!\!
 -{\vec Q}^2 \;{\omega}_{c}\; {\nu \over 2\pi}\;
                     ({{\vec Q}^2\over 2 {B_{\rm eff}}})^{m-3} \;
     {{4 m^2 (m-1)^2} \over {(m-1)!}} \;
                        \Big( 4 {{(m-1)}\over (m-2)}+
                        {2M {\tilde V}(0) \over 2\pi}\Big)^{-2}
\nonumber\\
&&
\label{eq:r23}
\end{eqnarray}
The above results are valid only if the pair potential ${\tilde V({\vec Q})}$,
has a gradient expansion in powers of $\vec Q$, {\it {\it i.~e.\/}},
for short range interactions.
${\tilde V}(0)$ stands for the leading order term in that expansion.

If the pair potential has the Coulomb form, {\it {\it i.~e.\/}},
${\tilde V}({\vec Q}) = {2\pi{q^2} \over |{\vec Q}|}$ in two spacial
dimensions, both, the dispersion relations with zero-momentum gap at
the cyclotron frequency and their residues get modified. The
expressions  valid in this case are, for any allowed value of $m$
\begin{equation}
 {\omega_{+}}(\vec Q) = \Big [{\omega _{c}}^2 +
                          {|{\vec Q}|\over 2 {B_{\rm eff}}} \;
                         {\bar \omega}_{c}^2 \; {2M{q^2}}
                          \Big ] ^{1\over2}
\label{eq:r24}
\end{equation}
with the same residue given by Eq.~\ref{eq:r21}, and
\begin{equation}
{\omega_{-}}(\vec Q) = \Big[ {\omega _{c}}^2 -
                        {|{\vec Q}|^{2m-3}\over ({2 B_{\rm eff}})^{m-2}} \;
                        {\bar \omega}_{c}^2
    {{4 m^2 (m-1)^2} \over {2M{q^2}(m-1)!}}\;     \Big] ^{1\over2}
\label{eq:r25}
\end{equation}
with residue
\begin{equation}
Res(K_{00},\omega_{-}({\vec Q})) = -{\vec Q}^2 \;
                                {\omega}_{c}\; {\nu \over 2\pi}\;
           {|{\vec Q}|^{2(m-2)}\over {(2{B_{\rm eff}})}^{m-3}} \;
        \times {{4 m^2 (m-1)^2} \over {(2M{q^2})^2(m-1)!}}
\label{eq:r26}
\end{equation}

In this section we have found the spectrum of collective excitations for
some values of the filling fraction. Our results are a generalization of
the work of Kallin and Halperin \cite{kallin} who studied the spectrum
of collective modes for the {\it integer} quantum Hall effect
within the RPA. We find a family of collective modes with dispersion
relations whose zero-momentum gap is $k{\bar \omega}_{c}$, where $k$ is an
integer number different from $1$ and $m$. When $k=m$,
{\it {\it i.~e.\/}}, the zero-momentum  gap is the
cyclotron frequency, there is a splitting in the dispersion relation for
finite wavevector. One of these solutions, $\omega_{-}$, has
negative slope for small values of $\vec Q$. Therefore, there must be a roton
minimum at some finite value of the wavevector. Since our results are
accurate only for small $\vec Q$, our dispersion curves do not apply close to
the roton minimum. Nevertheless, this mode is expected to become damped
due to non-quadratic interactions among the collective modes. On the other
hand, the collective mode with lowest energy, which has $k=2$, is stable
(at least for reasonably small wavevectors) and, at small wavectors, it
disperses downwards in energy. This behavior suggests that there should be a
magnetoroton minimum for this mode. This result is consistent with
the work of Girvin {\it et al.} \cite{girvin2}, where the authors used
the single mode approximation to obtain the lowest collective mode in
the lowest Landau level (the {\it intra} mode) for the states in the Laughlin
sequence.
The generalization of these results to other states of Jain's
hierarchy ($p \neq 1$) has been given in reference \cite{lopez3}.
Our results coincide, in the small momentum limit, with the ones
obtained by Simon {\it et al.} \cite{simonhalperin} more recently.
The splitting of the cyclotron mode for $\nu=1/3$ is a little
puzzling. It only happens
for $\nu=1/3$ and for short range interactions. In all other cases,
only the residue for one of the two cyclotron modes is proportional
to ${\vec Q}^2$. Standard lore has it that Kohn's theorem demands that there
should be one and only one mode converging to the cyclotron
frequency as ${\vec Q}^2 \to 0$ with residue proportional to ${\vec Q}^2$.
Zhang has emphasized this point recently \cite{zhang1}.
It is generally assumed that  Kohn's theorem is valid even at non-zero
wavevectors and that it requires the existence of only one mode with
residue proportional to ${\vec Q}^2$ converging to $\omega_c$. However,
at non-zero wave vectors, these arguments make the unstated
assumption of the analyticity of the current operators on the
wavevectors. While this may well be correct, it is an additional
assumption  and it does deserve closer scrutiny. The results from our
theory
do indeed predict the existence of only one mode {\it at} $\omega_c$
with residue proportional to ${\vec Q}^2$, which is the statement of
Kohn's theorem. And, also, for all filling
fractions and for all pair potentials (except $\nu=1/3$ and short range
interactions) we do find only one mode with residue ${\vec Q}^2$ even at
non-zero wavevectors. The case $\nu=1/3$ and short range interactions
appears to be exceptional in that we find two modes which coalesce at
the cyclotron frequency as ${\vec Q}^2 \to 0$. But both of these modes have
residue proportional to ${\vec Q}^2$ , with different amplitude, and
together they satisfy the sum rule \cite{lopez1}.
While it is possible that the non-gaussian corrections may change
this result since, in a sense, these are subleading pieces in ${\vec
Q}^2$, these non-gaussian corrections are expected to be very small
at small wavevectors.

We close this section with a few comments on the validity of this spectrum
of collective modes beyond the semiclassical approximation.
Primarily we have to consider the physics at moderately large wavevectors
and the (expected) effects of non-gaussian corrections. At the gaussian
(RPA) level we found a family of collective modes which, for sufficiently
small momentum, are infinitely long lived ({\it {\it i.~e.\/}}, the response
functions
have delta-function sharp poles at their location). These modes represent
charge-neutral bound states. It is in principle possible that,
for $\vec Q$ sufficiently large, these modes should become damped.
The threshold should occur when the energy of the collective mode becomes
equal to the energy necessary to create the lowest available
two-particle state: a quasiparticle-quasihole pair. In the AFA, the energy
of a pair is equal to ${\bar \omega}_c$. Gaussian fluctuations are expected
to renormalize this energy  and to give it a momentum dependence.
This is in principle calculable with the methods
described here \cite{simonhalperin}.
Non-gaussian corrections to the RPA are also expected to give a finite
width to (presumably) all the collective modes but the lowest one. This
is so because the corrections to the semiclassical
approximation are due to effective vertices (due to virtual
quasiparticle-quasihole pairs) which couple the various collective modes
and, thus, induce the higher energy modes to decay down into the lower
modes. However, by gauge invariance, these vertices have a momentum dependence
and should vanish as ${\vec Q} \to 0$. Thus, the width of the higher
energy modes goes to zero as ${\vec Q} \to 0$ and these modes only become
sharp at ${\vec Q}=0$. But at ${\vec Q}=0$ the only accessible mode is the
cyclotron mode (the other modes have a vanishingly small spectral
weight). These arguments strongly suggest that the only truly sharp mode,
{\it at} ${\vec Q}=0$, is the cyclotron mode, which is required to be stable
by Kohn's theorem \cite{kohn}. Since the modes with zero momentum gap at
$k{\bar \omega_{c}}$, $k \geq 3$, are not the collective modes with lowest
energy, it is possible that at finite wavevectors they may also decay into
the collective mode with lowest energy ( the mode with
$k=2$, which has a gap at ${\bar \omega}_c$).

\subsection{Optical properties and experiments}
\label{subsec:optical}

In this section we discuss the experimental consequences of
the results that we have just derived.
The density correlation function can be probed by optical absorption and by
Raman-scattering experiments.

In the first case, the optical absorption is
proportional to the imaginary part of the density correlation function.
We predict that there will be absorbtion peaks at a discrete set of frequencies
which, for ${\vec Q}\to 0$ converge to
$\omega = k {\bar \omega }_{c}$, where $k$ is an integer number greater than
two. Since the spectral weight of these modes vanishes as ${\vec Q}\to
0$, the associated absorption peaks are,  for a strictly
translationally invariant system,  only observable at non-zero momentum.

In the case of the Raman scattering, the geometry must be such that there is a
component of the incident light wavevector in the plane of the sample.
The Raman spectrum, $I(\omega)$, is also proportional to the imaginary part
of the density correlation function \cite{klein}.

We have seen that in the limit $|{\vec Q}|\ll {\ell}^{-1}$, where $\ell$ is
the magnetic length, most of the weight of $K_{00}(\omega,{\vec Q})$
is in one of the cyclotron modes. The pole in $K_{00}(\omega,{\vec
Q})$ for the lowest excitation frequency, $\omega _k$ with $k=2$,
has a residue which is proportional to $|{\vec Q}|^4$, {\it {\it
i.~e.\/}}, it is smaller by a factor of $|{\vec Q}|^2$ than the
residue  at the highest weighted mode at the cyclotron frequency.

We have also found that there is a splitting in the cyclotron modes.
If the pair potential has a gradient expansion in $|{\vec Q}|$,
{\it {\it i.~e.\/}}, short range interaction, the pole at $\omega_{-}$
(Eq.~\ref{eq:r22}), has a residue that is smaller by a factor of
$|{\vec Q}|^{2(m-3)}$ than the residue of $\omega _{+}$
(Eq.~\ref{eq:r20}). The relative Raman
intensity, ${I(\omega_{+})\over I(\omega_{-})}$, is proportional to
$({{2B_{\rm eff}} \over {\vec Q}^{2}})^{(m-3)}$ which is a big
number within our approximation.
If the filling fraction is $\nu ={1\over 3}$, both modes have the same
${\vec Q}^2$ dependence in their spectral weight, but the relative
intensity is $\approx 2.5 $ provided that ${\tilde V}(0)=0$.
Except for $\nu={1\over 3}$, the splitting between the two modes at the
cyclotron frequency satisfies, at leading order in $|{\vec Q}|$,
$\triangle \omega ^2 = \omega _{+}^{2} - \omega _{-}^{2}
= \omega _{+}^{2} - \omega _{c}^{2}$, which is proportional to $|{\vec Q}|^2$.
Up to this order, experimentally one should observe one mode dispersing as
$\omega _{+}$ (Eq.~\ref{eq:r20}), and the other as $\omega =\omega _{c}$.
For $\nu = {1\over 3}$ the splitting is also proportional to $|{\vec Q}|^2$.
In this case one should observe both modes ($\omega _{+}$ and
$\omega _{-}$, Eq.~\ref{eq:r17}), but with different intensities.

If the pair potential has the Coulomb form, the residue of $\omega_{-}$
(Eq.~\ref{eq:r26}) is smaller by a factor of
$|{\vec Q}|^{2(m-2)}$ than the residue of $\omega _{+}$ (Eq.~\ref{eq:r24}),
and this is valid
for all the values of the filling fraction that we have studied.
The splitting between these two modes satisfies, at leading order in
$|{\vec Q}|$, $\triangle \omega ^2 = \omega _{+}^{2} - \omega _{c}^{2}$,
which is proportional to $|{\vec Q}|$. For $\nu$ different from $1\over 3$,
the relative intensity between the two modes is proportional to
$({{2B_{\rm eff}}/ {\vec Q}^{2}})^{(m-3)}{M{\tilde V}(\vec Q)}$, which is
bigger than $1$ within our approximation. For $\nu = {1\over 3}$,
the relative intensity is proportional to ${M{\tilde V}(\vec Q)}$. This factor
can be written in terms of the magnetic length and the cyclotron energy
as follows $ {{\tilde V}(\vec Q) / {\ell} \over \omega_{c}}$. Since
our approximation is only valid in the limit
${1\over |{\vec Q}|}\gg {\ell}$, the numerator satisfies
${\tilde V}(\vec Q) / {\ell} \gg { 2\pi {q^2} \over {\ell}}$. The second
term in this inequality is the Coulomb energy at the magnetic length, which is
typically of the same order of magnitude than the cyclotron energy.
Therefore,
${{\tilde V}(\vec Q) / {\ell} \over \omega_{c}} \gg { 2\pi {q^2} / {\ell}
\over \omega_{c}} \approx 1$. In other words, the relative intensity for
$\nu ={1\over 3}$ is also bigger than one.

\subsection{Saturation of the $f$-sum rule}
\label{subsec:f-sum}

We show now that the long wavelength form of $K_{00}$, found at this
semiclassical level, saturates the $f$-sum rule. This result implies
that the non-gaussian corrections do not contribute at very small
momentum. In Section \ref{sec:wave} we will use this
result to show that the absolute value squared wave function of all the
(incompressible) liquid states has the Laughlin form at very long
distances, in the thermodynamic limit.

The retarded density and current correlation functions of this theory are, by
definition
\begin{equation}
D^R_{\mu \nu}(x,y)  = -i \theta (x_{0}-y_{0})
 \times <G| [J_{\mu}(x),J_{\nu}(y)] |G>
\label{eq:r36}
\end{equation}
where $J_{\mu}$ ($\mu =0,1,2$) are the conserved  currents of the theory
defined by Eq.~\ref{eq:c12}, and $|G>$ is the ground state of the system.
Using this definition and the commutation relations between the currents,
one can derive the $f$-sum rule for the retarded density correlation
function $D^R_{00}$. In units in which $e=c=\hbar =1$, it states that
\begin{equation}
\int _{-\infty}^{\infty } \; {d\omega \over 2\pi} \; i \omega
       D^R_{00}(\omega,{\vec Q}) = {{\bar \rho} \over M}{\vec Q}^2
\label{eq:r37}
\end{equation}
The polarization tensor $K_{00}$ and the density
correlation function $D_{00}$ are simply related by $K_{00}=-D_{00}$.
~From Eq.~\ref{eq:r12} and ~\ref{eq:r13} we see that the leading
order term in ${\vec Q}^2$ of the zero-zero component of the
electromagnetic response is given by
\begin{equation}
K_{00} = - {{\bar \rho}\over M}\; {{\vec Q}^2 \over {{\omega }^2 -
    {\omega }^2_c} + i \epsilon}
\label{eq:r39}
\end{equation}
where we have used that ${{\bar \rho}\over B} = {\nu \over 2\pi}$.
Thus, we see that the leading
order term of $K_{00}$ saturates the $f$-sum rule, Eq.~\ref{eq:r37}.

It is important to remark that the coefficient of the leading order term of
$K_{00}$ can not be renormalized by higher order terms in the gradient
expansion, nor in the semiclassical expansion. In the case
of the gradient expansion, it is clear that higher order terms have higher
order powers of ${\vec Q}^2$, and then, do not modify the leading order term.
In the case of the corrections to $K_{00}$ originating in higher order terms
in the semiclassical expansion, they also come with higher order powers of
${\vec Q}^2$. The reason of that is essentially the gauge invariance of the
system. This implies that the higher order correlation functions must be
transverse in real space, or equivalently they have higher order powers of
${\vec Q}^2$ in momentum space.
Being higher order terms in the ${\vec Q}^2$ expansion they can not change
the leading order term.

As we have already mentioned, these results hold for any model Hamiltonian
for the two-dimensional electron gas (2DEG) with reasonably local interactions,
 {\it {\it i.~e.\/}}, with pair interactions that obey
${\vec Q}^2 {\tilde V}(Q)  \rightarrow 0$ as ${\vec Q}^2 \rightarrow 0$.

\subsection{Hall Conductance, Effective Low Energy Action and
Quantum Numbers}
\label{subsec:hallsingle}

We show now that, already within our approximations, this state does
exhibit the Fractional Hall Effect with the exact value of the Hall
conductance.
In order to do so, we will calculate the Hall conductance of the whole system.
Since we are only interested in the leading long-distance behavior,
it is sufficient to keep only  those terms in Eq.~\ref{eq:r11}
which have the smallest number of derivatives, or
in momentum space, the smallest number of powers of $\vec Q$.
Therefore, from Eq.~\ref{eq:r11} and ~\ref{eq:r12}, we see that the leading
long distance behavior ({\it {\it i.~e.\/}}, small momentum)  of
the effective action for the  electromagnetic field is governed by
the Chern-Simons term. In this limit Eq.~\ref{eq:r11} turns out to be
\begin{equation}
S_{\rm eff}^{\rm em} ({\tilde A}_{\mu}) \approx -{i\over 2}
                       \int {d^{2}Q d\omega \over (2\pi)^3}
                       {\tilde A}_{\mu}(-\omega ,-{\vec Q})
            {K_1}(\omega ,{\vec Q}) {\epsilon _{\mu\nu\lambda}}
                       { Q^{\lambda}}{\tilde A}_{\nu}(\omega ,{\vec Q})
\label{eq:r40}
\end{equation}
where $Q^{0}=\omega$ and $Q^{i}=-Q_{i}$.

To  study the Hall response of the system, we will now consider the
limit of small $\omega$ and
small $\vec Q$.
We have checked that in this theory the two limits commute
\begin{equation}
{\lim _{{\vec Q} \rightarrow 0}}\; {\lim _{\omega \rightarrow 0}}
{ K_1}(\omega,{\vec Q}) ={{\Pi_{1}(0,0)}\over{\theta + \Pi_{1}(0,0)}}
\label{eq:r41}
\end{equation}
\begin{equation}
{\lim _{\omega \rightarrow 0}}\; {\lim _{{\vec Q} \rightarrow 0}}
{ K_1}(\omega,{\vec Q}) ={{\Pi_{1}(0,0)}\over{\theta + \Pi_{1}(0,0)}}
\label{eq:r42}
\end{equation}
This is a consequence of the incompressibility of the ground state.
Since $\theta = {1\over 2\pi \;2s}$ and $\Pi_{1}(0,0)=
{p\over 2\pi}$
\begin{equation}
K_{1}\rightarrow {\nu \over 2\pi} \equiv \theta_{\rm eff}
\label{eq:r43}
\end{equation}
where $\nu$ is the filling fraction.
The electromagnetic current
$J_{\mu}$ induced in the system is obtained by differentiating the
effective action $S_{\rm eff} ({\tilde A}_{\mu})$ with respect
to the electromagnetic vector potential. The current is
$J_{\mu}={\theta_{\rm eff}\over 2} {\epsilon}_{\mu\nu \lambda} {\tilde F}^{\nu
\lambda}$.
Thus, if a weak external electric field ${\tilde E}_j$ is applied, the
induced current is
$J_k=\theta_{\rm eff} {\epsilon}_{lk} {\tilde E}_l $.
We can then identify the coefficient $\theta_{\rm eff}$ with the {\it
actual} Hall conductance of the system ${\sigma}_{xy}$ and get
\begin{equation}
{\sigma}_{xy}\equiv \theta_{\rm eff}={\nu\over 2 \pi}
\label{eq:r44}
\end{equation}
which is a {\it fractional} multiple of ${e^2\over h}$ (in units in
which $e=\hbar=1$). Thus, the uniform states exhibit a Fractional
Quantum Hall effect with the correct value of the Hall conductance.

We can also derive these results by considering an effective action for
the low energy excitations.  At the gaussian level, the effective
action is obtained by keeping only those terms in Eq.~\ref{eq:c40} 
which have the smallest number of derivatives and  we can then 
approximate $S_{\rm eff}$ by
\begin{equation}
S_{\rm eff} \approx {\sigma_{xy}^0\over 4} S_{\rm CS}({\tilde a}_{\mu})+
{\theta \over 4} S_{\rm CS}({\tilde a}_{\mu}-{\tilde A}_{\mu}).
\label{eq:c49}
\end{equation}
where the induced Hall conductance $\sigma_{xy}^0=\pm {p\over {2\pi}}$.
Although we have obtained this result within the gaussian approximation,
it should be stressed that
$\sigma_{xy}^0$ does not get corrected to any order in perturbation theory.
This is so because, at least for a system with a gap and in the
thermodynamic limit, $\sigma_{xy}^0$ is determined by a topological
invariant \cite{niu,kohmoto}, the First Chern Character ${\cal C}$. In the
problem at hand, we have ${\cal C}=p$. Thus, the Hall conductance that we
will calculate below is actually exact.

Upon integrating over the statistical vector potentials, we find that
the effective action for the electromagnetic fields is
\begin{equation}
S_{\rm eff}^{\rm em} ({\tilde A}_{\mu}) \approx {\theta_{\rm eff} \over
4} S_{\rm CS}({\tilde A}_{\mu}).
\label{eq:c50}
\end{equation}
where $\theta_{\rm eff}$ is given by
\begin{equation}
{1\over \theta_{\rm eff}}={1\over \theta}+{1\over \sigma_{xy}^0}.
\label{eq:c51}
\end{equation}
{\it i.~e.\/},  the Chern-Simons coupling constants are added ``in parallel".

The values of $\theta$ and $\sigma_{xy}^0$ determined above yield the result
\begin{equation}
\theta_{\rm eff}={\nu \over 2 \pi}
\label{eq:c52}
\end{equation}
where $\nu$ is the filling fraction. As before, we obtain the value
of the Hall conductance from the electromagnetic current
$J_{\mu}$ induced in the system. The result is
$\sigma_{xy}\equiv \theta_{\rm eff}={\nu\over 2 \pi}$.
Notice that the coefficient $\sigma_{xy}^{\rm eff}$ of
the effective action represents the integer Hall effect of the bound
states and it is different from the true Hall conductance $\sigma_{xy}$.

Finally, we discuss the charge and statistics of the topological excitations,
the quasiparticles and quasiholes. We recall that, at the level of the
Average Field Approximation, the single particle excitations are {\it
fermions}, single (charge one) composite fermions. However, the
fluctuations of the statistical gauge fields modify this picture in an
essential way. For instance, if we want to compute the effective charge
of these excitations, we need to find the exact low energy behavior of
the composite fermion propagator. Since, even at the level of the AFA,
these single particle states have a finite energy gap, the Feynman path
integral form of their propagator is dominated by smooth paths $\Gamma$.
The fermions couple to both, the statistical and electromagnetic fields
as if they had charge one. It was shown in reference \cite{book}
that the leading effect of the fluctuations of the
gauge fields can be found by computing the expectation value of a Wilson
loop operator using the effective action of Eq.~\ref{eq:c49}. More
explicitly, we find
\begin{equation}
\langle e^{i\int_\Gamma \; dx_\mu (a^\mu(x)+A^\mu(x))}\rangle_{\rm eff}=
e^{i\gamma(\Gamma)} \;\langle e^{i\int_\Gamma \; dx_\mu \; {\tilde
a}^\mu(x)}\rangle_{\rm eff}
\label{eq:single1}
\end{equation}
Here  $\gamma(\Gamma)$ is the  Aharonov-Bohm phase factor
\begin{equation}
\gamma(\Gamma)= {\frac{2\pi}{2sp \pm 1}}{\frac{\Phi_0(\Gamma)}{\phi_0}}
\label{eq:flux}
\end{equation}
where $\Phi_0(\Gamma)=BL(\Gamma)^2$ is the flux through the area
enclosed by $\Gamma$ and $\phi_0$ is the flux quantum. 
On the l.h.s. of this equation $a^{\mu}$ is the complete Chern-Simons
gauge field, {\it i.e.}, the sum of the solution of the SPE plus the 
fluctuating part ${\tilde a}^{\mu}$. Eq.~\ref{eq:flux}
tells us that the {\it effective charge} of this excitation is $q_{\rm
eff}={\frac{e}{2sp \pm 1}}$. The additional phase factor can, in
principle, give rise to a fractional spin but it is not possible to
determine its value (or even if it exists!) without making contact with
the microscopic theory.

The statistics of these excitations can be computed using a similar type
of reasoning. Now we need to look at a two-particle propagator and,
consequently, we need to compute expectation values of two Wilson loops.
In section \ref{sec:csfqhe} we gave a prescription for the
computation of the expectation values that we need here. The effective
value of $\theta$ in Eq.~\ref{eq:wl}
now is ${\bar \theta}=\theta + \sigma_{xy}^0$. Thus, the
statistical angle $\delta$ (measured from fermions) is
\begin{equation}
\delta={\frac{1}{2{\bar \theta}}}=\pi {\frac{2s}{2sp\pm 1}}
\label{eq:delta}
\end{equation}
Hence, at long distances and at low energies, the composite fermions are
actually fractionally charged anyons. This is the well know result of
Arovas, Scrhieffer and Wilczek \cite{ASW}.

\section{Fermionic Chern-Simons theory for the Fractional Quantum Hall
Effect in bilayers and partially polarized systems}
\label{sec:csbil}

If one allows for the presence of new degrees of freedom, a richer variety
of states can be found. The two obvious possibilities that one can
consider
are systems in which the electronic spin is not frozen by the Zeeman
energy,
and systems in which two or more layers of 2DES are coupled together.
For instance, the experimentally observed $\nu ={5\over 2}$ state
\cite{willet}, has been explained theoretically by Haldane and Rezayi
\cite{hr} using the fact that the system is not spin-polarized.

Due to continuing advances in material-growth techniques, it has been
possible
to fabricate high-quality multiple 2DE layers in close proximity. In these
systems the layer index is the new degree of freedom, and the interplay
between the intralayer and the interlayer Coulomb interactions gives
rise to
very interesting physics. In particular, this competition can explain
\cite{mpb} the experimental observation \cite{exp2} of the destruction or
weakening of the IQHE at odd filling fractions.
 Another interesting case is the one of the $\nu = {1\over 2}$ state.
In single-layer systems, even though many transport anomalies have been
reported, there is no evidence of FQHE. On the other hand, this is a
well observed \cite{unmed} FQHE state in double-layer systems.

There are two energy scales that play a very important role in bilayers.
One is the potential energy between the electrons in different layers, and
the other one is the tunneling amplitude between layers.
In the case in which
tunneling between the layers may be neglected, the number of particles
in each layer is conserved. The system
has an effective $U(1) \otimes U(1)$ symmetry reflecting the separate
conservation of charge in each layer. Tunneling processes break this
symmetry down to a global $U(1)$ invariance. In the next subsection
we will describe the $U(1) \otimes U(1)$ fermionic Chern-Simons theory for
bilayers which ignores interlayer tunneling effects. In a subsequent
section we present an $SU(2) \otimes U(1)$ fermionic Chern-Simons theory.
This theory describes the physics of spin as well as
tunneling effects in bilayers.

\subsection{Fermionic Chern-Simons theory for double layer FQHE systems}
\label{subsec:CS}

The action for a double-layer 2DES
in the presence of an external uniform magnetic field $B$ perpendicular to it
is given by
\begin{eqnarray}
{\cal S}&=&\int  d^3 z \; \sum _{\alpha}\; \left\{ \psi^*_{\alpha}(z)
[i D_0 +\mu_{\alpha}]
\psi_{\alpha}(z)-{1 \over2 M}|{\vec D}\psi_{\alpha}(z)|^2 \right\}
\nonumber \\
&&-{1\over 2} \int  d^3z \int d^3z'\; \sum _{\alpha ,\beta}\
(|\psi_{\alpha}(z)|^2-{\bar\rho}_{\alpha})
   V_{\alpha \beta}(|{\vec z}-{\vec z}'|)
(|\psi_{\beta }(z')|^2-{\bar\rho}_{\beta}) \nonumber \\
\label{eq:ese1}
\end{eqnarray}
where the indices $\alpha =1,2$ and $\beta =1,2$ label the layers,
${\bar \rho}_{\alpha}$ is the average particle density in the layer
$\alpha$, $\psi(z)_{\alpha}$ is a second
quantized Fermi field, $\mu_{\alpha}$ is the chemical potential and
$D_{\mu}$
is the
covariant derivative which couples the fermions to the external
electromagnetic
field $A_{\mu}$. In what follows we
will assume that the pair potential has either the Coulomb form, {\it i.~e.\/},
\begin{equation}
V(|\vec r|)_{\alpha\beta} =  {q^2 \over {\sqrt{{\vec r\;}^2 +
{\vec d \;}^2 (1-
\delta_{\alpha\beta}))}}}
\label{eq:int}
\end{equation}
(with $d$ the interlayer separation), or that it represents a short range
interaction such that in momentum space it satisfies that
${V}_{\alpha \beta}(\vec Q){\vec Q}^2$ vanishes at zero momentum.
This includes the case of ultralocal potentials ({\it i.~e.\/}, with
a range smaller or of the same order as the cyclotron length $\ell$),
in which case we can set ${\tilde V}(0) = 0$,
or short range potentials with a range longer than $\ell$ such as a
Yukawa interaction.

Using a generalization of the methods of Section \ref{sec:csfqhe}, it is
possible to show~\cite{lopez5} that this system is equivalent to a
system of interacting electrons coupled to an additional statistical
vector potential
$a^{\mu}_{\alpha}$ ($\mu=0,1,2$) whose dynamics is governed by the
Chern-Simons action
\begin{equation}
{\cal S}_{\rm cs}= \sum _{\alpha\beta}\; {\kappa _{\alpha \beta}\over 2}
\int d^3x\;
\epsilon^{\mu \nu \lambda} {a}_{\mu}^{\alpha} \partial_{\nu}
{a}_{\lambda}^{\beta}
\label{eq:cs}
\end{equation}
provided that the CS coupling constant matrix satisfies
\begin{eqnarray}
\kappa^{\alpha \beta}={1\over {2\pi (4 s_1 s_2 - n^2)}}
\left(
\begin{array}{cc}
2 s_2  & -n \\
-n & 2 s_1
\end{array}
\right)
\label{eq:Kalfabeta}
\end{eqnarray}
where $s_1$, $s_2$ and $n$ are arbitrary integers. In Eq.~\ref{eq:cs}
$x_0,x_1 \;{\rm and}\; x_2$ represent the time and the space
coordinates of the electrons  respectively.
In the equivalent theory the covariant derivative is given by
\begin{equation}
D_{\mu}^{\alpha}=\partial_{\mu}+i{e\over c}A_{\mu}+i a_{\mu}^{\alpha}
\label{eq:dcov}
\end{equation}
and it couples the fermions to the statistical gauge fields
($a_{\mu}^{\alpha}$),
and to the external electromagnetic field ($A_{\mu}$). Notice that the
theory has now a $U(1) \otimes U(1)$ gauge invariance.

The Chern-Simons coupling constant matrix $\kappa^{\alpha \beta}$ of
Eq.~\ref{eq:Kalfabeta} must be non-singular. For the case of
a singular matrix, it is better to arrange the gauge fields in linear
combinations $a_{\pm}^\mu$ which diagonalize the coupling matrix.
In general this situation arises only if the matrix elements satisfy
$\kappa_{11}=\kappa_{22}\equiv \kappa$ and
$\kappa_{12}=\kappa_{21}=\kappa'$ and the matrix is singular if
$\kappa=\kappa'$. In this case
the gauge field $a_-^\mu$  acts like a local
constraint on the allowed states demanding that
$\rho_1({\vec x},t)=\rho_2({\vec x},t)$.
In other words, the allowed states consist of {\it pairs} of
fermions, one from each layer. If we think of the layer index as spin,
these states are local spin singlets and the allowed states are spinless
{\it bosons} of charge $2$ (with zero transversal spatial extent). Thus,
for a singular coupling matrix, this theory effectively
describes a FQHE of a single layer of charge $2$ bosons. While this is
prefectly correct, it is clear that this is an extreme case
and that this approach is uncapable of describing a singlet FQH state not
made of local singlets. To describe a state of this sort within a
Chern-Simons approach it is required to attach separate fluxes to each
species of fermions in a manner consistent to the full $SU(2)$ symmetry
of spin. To describe this situation correctly it is necessary to
use a non-abelian approach, which we present in  Section~ \ref{subsec:su2}.

There is another limit of physical interest. Clearly, in a
bilayer system (or for a single layer with spin) it should also
be possible to attach the same flux (using only one gauge field)
to the fermions of both layers.
This situation is achieved by taking the limit $\kappa-\kappa' \to \infty$
(which kills the field $a_-^\mu$)
while holding ${\frac{1}{2}} (\kappa+\kappa')$ finite.
This is the conventional scheme used to describe partially
polarized FQH states \cite{skkr}.

The Chern-Simons action implies a constraint for the particle density
$j_0^{\alpha}({\vec x})$ and the statistical flux
${\cal B}^{\alpha}={\epsilon }_{ij} {\partial}_{i} a_{j}^{\alpha}$,
given by
\begin{equation}
j_0^{\alpha}({\vec x}) + \kappa _{\alpha \beta}
{\cal B}_{\beta}(\vec x) =0
\label{eq:cons23}
\end{equation}
(from now on we assume that repeated indices are contracted).

This relation states that the electrons in plane ${\alpha}$ coupled to
statistical gauge fields with Chern-Simons coupling constant given by
Eq.~\ref{eq:Kalfabeta}, see a statistical flux per particle of
$2\pi 2s_{\alpha}$ for the particles in their own plane , and
a statistical flux per particle of $2\pi n$ for the particles in the
opposite
plane. (Notice that in units in which $e=c=\hbar=1$, the flux quantum is
equal to $2\pi$).
If the coefficient of the Chern-Simons
term is chosen with the above prescription, all the physical amplitudes
calculated in this theory are
identical to the amplitudes calculated in the standard theory, in which
the
Chern-Simons field is absent. Of course, this is true provided that the
dynamics of the statistical gauge fields is fully taken into account
exactly.

Here we  take into account the dynamics of the Chern-Simons gauge
fields in a semiclassical expansion, which is a sequence of well
controlled approximations. In practice, we will only consider the
leading and next-to-leading order in the semiclassical approximation.

Using the constraint enforced by the Chern-Simons action, the
interaction term of the action Eq.~\ref{eq:ese1} becomes
\begin{equation}
{\cal S}_{int}= - {1\over 2} \int  d^3z \int d^3z'\;
(\kappa ^{\alpha \delta} {\cal B}_{\delta}(z)-
{\bar\rho}_{\alpha}) \;V_{\alpha \beta}(|{\vec z}-{\vec z}'|)\;
(\kappa ^{\beta \gamma} {\cal B}_{\gamma}(z')-{\bar\rho}_{\beta})
\label{eq:eseint}
\end{equation}

The quantum partition function for this problem is, at zero temperature
\begin{equation}
{\cal Z}[A_{\mu}]=\int {\cal D} \psi^* {\cal D} \psi {\cal D}
a_{\mu}^{\alpha} \; \exp (i S(\psi^*,\psi,a_{\mu}^{\alpha},A_{\mu}))
\label{eq:pf}
\end{equation}
Since the action is quadratic in the fermions, they can be integrated out.
The effective action (${\cal S}_{\rm eff}$) is given by
(in units in which $e=c=\hbar=1$)
\begin{equation}
{\cal S}_{\rm eff}= -i \sum _{\alpha} {\rm tr} \; {\ln} \left\{
i D_0^{\alpha} +\mu_{\alpha}+
{1 \over2 M}({\vec D}^{\alpha})^2 \right\}
+ {\cal S}_{\rm cs} ( a_{\mu}^{\alpha} - {\tilde A}_{\mu}^{\alpha})
+ {\cal S}_{\rm eff}^{\rm int} ( a_{\mu}^{\alpha} -
{\tilde A}_{\mu}^{\alpha})
\label{eq:ese2}
\end{equation}
where
\begin{eqnarray}
{\cal S}_{\rm eff}^{\rm int} ( a_{\mu}^{\alpha} - {\tilde A}_{\mu}^{\alpha})
=&&
-{1\over 2} \int  d^3z \int d^3z'\;(\kappa ^{\alpha \delta}
({\cal B}_{\delta}(z)-{\tilde B}_{\delta}(z))- {\bar\rho}_{\alpha})
\nonumber
\\
&& V_{\alpha \beta}(|{\vec z}-{\vec z}'|)
(\kappa ^{\beta \gamma} ({\cal B}_{\gamma}(z')-{\tilde B}_{\gamma}(z'))
-{\bar\rho}_{\beta})
\label{eq:esein2}
\end{eqnarray}
Here we have written the external electromagnetic field as
a sum of two terms, one representing the uniform magnetic
field $B$, and a small fluctuating term ${\tilde A}_{\mu}^{\alpha}$
whose average
vanishes everywhere. The latter will be used to probe the electromagnetic
response of the system. Notice that we have used the invariance of the
measure ${\cal D} a_{\mu}^{\alpha}$ with respects to shifts, to move
${\tilde A}_{\mu}^{\alpha}$ out of the covariant derivatives and into
the Chern-Simons
and the interaction terms of the effective action (Eq.~\ref{eq:ese2}).

The path integral $\cal Z$ can be approximated by expanding its degrees of
freedom in powers of the fluctuations, around stationary configurations of
${\cal S}_{eff}$.
This requirement yields the SPE or classical equations of motion.
\begin{eqnarray}
< j_0^{\alpha} (z)>_{F}&=& - {\kappa}_{\alpha\beta}
[< {\cal B}^{\beta}(z)>-
{\tilde B}^{\beta}(z)> ] \nonumber\\
< j_k^{\alpha} (z)>_{F}&=& - {\kappa}_{\alpha\beta} \epsilon_{kl}
[< {\cal E}^{\beta}_{l}(z)>-{\tilde E}^{\beta}_{l}(z)>] \nonumber \\
&& +  \epsilon_{kl} \partial_{l}^{z} \int d^3 z' \;
\kappa^{\alpha\epsilon}
V_{\epsilon\delta}(z,z') [-{\kappa}^{\delta\gamma}<{\cal B}^{\gamma}-
{\tilde B}^{\gamma}>(z')- {\bar \rho}^{\delta}]
\nonumber \\&&
\label{eq:sp}
\end{eqnarray}
Just as in the case of the single layer problem, these equations have many
possible solutions.
The uniform solutions of Eq.~\ref{eq:sp} represent the FQH fluid states
of the bilayer system at the mean field level.
In this state, the electrons in the layer $\alpha$ see a total flux
$B_{\rm eff}^{\alpha}$, equal to the
external magnetic flux partially screened by the average Chern-Simons flux,
{\it i.~e.\/} $B_{\rm eff}^{\alpha}=B+\langle {\cal B}^{\alpha}\rangle
=B-({\kappa ^{-1})^{\alpha\beta} {\bar \rho}_{\beta}}$.
It is easy to see that the uniform saddle-point state has a gap only if the
effective field $B_{\rm eff}^{\alpha}$ is such that the fermions in layer
$\alpha$ fill exactly an integer number $p_{\alpha}$ of the effective
Landau levels, {\it i.~e.\/}, those defined by $B_{\rm eff}^{\alpha}$.
In other words, the AFA to this theory yields a state with an energy gap if
the filling fractions of each layer satisfy
\begin{eqnarray}
\nu _{1} &=& {1 \over {\pm {1\over {p_{1}}}+ 2s_{1} + n {N_{2}\over N_{1}}}}
\nonumber \\
\nu _{2} &=& {1 \over {\pm {1\over {p_{2}}}+ 2s_{2} + n {N_{1}\over N_{2}}}}
\label{eq:fill}
\end{eqnarray}
where $N_{1}$ and $N_{2}$ are the number of particles in layers $1$ and $2$
respectively. The sign in front of $p_{1}$ and $p_{2}$ indicates if the
effective field is parallel or antiparallel to the external magnetic field.

Using the fact that $\nu = \nu_{1} + \nu_{2}$, and that the number of flux
quanta enclosed by each plane is the same, the filling fractions can be
written as follows
\begin{eqnarray}
\nu _{1} &=& {{n-(\pm {1\over p_{2}} + 2 s_{2})} \over
              { n^{2} - (\pm {1\over p_{1}} + 2 s_{1})(\pm {1\over p_{2}}
+ 2 s_{2})}}
\nonumber \\
\nu _{2} &=& {{n-(\pm {1\over p_{1}} + 2 s_{1})} \over
              { n^{2} - (\pm {1\over p_{1}} + 2 s_{1})(\pm {1\over p_{2}}
+ 2 s_{2})}}
\label{eq:fillfrac}
\end{eqnarray}
where $p_{1}$, $p_{2}$, $s_{1}$, $s_{2}$ and $n$ are arbitrary integers.
For the special case in which the two layers have the same occupancy,
$N_1=N_2$ and $\nu_1=\nu_2={\frac{\nu}{2}}$, the allowed fractions are
\begin{equation}
\nu(p,n,s)={\frac{2p}{(n+2s)p+1}}
\label{eq:simplerfraction}
\end{equation}
where $p$ is an arbitrary (positive or negative) integer.

The effective magnetic field can be written in terms of the external
magnetic field as
\begin{equation}
B_{\rm eff}^{\alpha}= B { \nu^{\alpha} \over  p^{\alpha}}
\label{eq:beff}
\end{equation}
For general values of these integers, the states whose filling fractions
are given by Eq.~ \ref{eq:fillfrac} have a gap at the mean field level
of this approximation. This ensures that the perturbative expansion is
meaningful.  If there is no gap for the excitations of the mean field
ground state, the perturbative expansion breaks down.
The breakdown is signalled by infrared divergencies at low temperatures.
Such is the case for the compressible or ``Fermi Liquid" states.

It is clear from Eq.~\ref{eq:fillfrac} that in the bilayer systems,
there are many possible choices of the numbers $p_{1}$, $p_{2}$,
$s_{1}$, $s_{2}$  and $n$, {\it i.~e.\/}, many different states,
which have the same filling fractions. These states are physically
different since, as we will see, the quantum numbers of their
excitation spectra are different. This is an example of the concept of
{\it topological order} that has been formulated by X.~G.~Wen~
\cite{wenorder}. More recently, Haldane \cite{haldane-stability} has
proposed an algebraic criterion to determine what states are
topologically stable. We will see that all the solutions with an
energy gap  for all excitations are topologically stable in the
sense of Haldane.

As a result of the presence of many solutions, the phase diagram for
bilayers is much richer than for
spin polarized electrons in a single layer. Which particular solution is
actually realized is determined by the microscopic dynamics of a
particular system and it requires a detailed calculation of the ground
state energy beyond the Average Field Approximation.
Experiments on spin polarized 2DEG bilayers \cite{exp2,unmed} at
filling fractions $\nu=1,1/2$ have
shown a non-trivial phase diagram with at least two phases: a
compressible phase in the regime in which the 2DEG's are well separated
and an incompressible phase when they are closer by.  However, as it was
emphasized by Wen and Zee \cite{wen}, the
incompressible states at these two filling fractions actually have quite
different properties. We will see now that the phase diagram can be
quite complex.

Let us consider some cases of special interest, in particular with
$N_1=N_2$. For example, at level one of the hierarchy, we choose
$p_{1}=p_{2}=1$, $\; 2s_{1}+1 = m_{1}$ and $\; 2s_{2}+1
= m_{2}$, and we obtain the so called $(m_1, m_2, n)$ states
\cite{wen,halperin1,wilczekd} whose filling fractions
are $\nu = {m_{1} + m_{2} - 2n \over {m_{1} m_{2} - n^2}}$. In
particular,
for $m_{1}=m_{2}=3$ and $n=1$, this is the $(3,3,1)$ state, whose filling
fraction is $\nu = {1\over 2}$.
For $p_{1}=p_{2}=1$, and $2s_{1}+1=2s_{1}+1=n=m$, we obtain the $(m, m, m)$
states \cite{wen,ezawa} whose filling fractions are $\nu = {1 \over m}$.

The states $(m_1, m_2, n)$ are the  equivalent of the Laughlin states for
bilayers and as such should be regarded as primary states. Hence, for
each $(m_1, m_2, n)$ primary state there is a hierarchy of states
labelled by the index $p$. For example, the primary state $(1,1,1)$,
defined by $p=1$, $s=0$ and $n=1$, has the hierarchy
$\nu_+(p,1,0)={\frac{2p}{p+1}}=1,{\frac{4}{3}},{\frac{3}{2}},{\frac{8}{
5}},\ldots, 2$. The first three states in this series have already
been seen in experiments \cite{hamilton}.
Similarly, the primary state $(3,3,1)$ (which has
$p=1$, $n=s=1$), has the hierarchy
$\nu_+(p,1,1)={\frac{2p}{3p+1}}={\frac{1}{2}},
{\frac{4}{7}},{\frac{3}{5}},\ldots, {\frac{2}{3}}$. Recall that
$\nu_\pm(p,n,s)$ is the {\it total} filling fraction of the bilayer and
that the individual filling fractions of each layer are $\nu/2$.

We can also consider the limit $p_{1}=p_{2}\equiv p \rightarrow \infty$.
In this case, if $s_{1}=s_{2}=s$, the state with filling fraction $\nu={1
\over m}$ will be obtained for all the values of $n$ and $s$ such that
$n+2s= 2m$.
Since $p \rightarrow \infty$, the effective field vanishes, and we find the
analogous of the compressible states for the single layer problem discussed
by Halperin et al \cite{hlr}.
All of these states are degenerate with the
$\nu ={1\over m}$ states mentioned above.

Clearly, it is always possible to construct a large family of, in
principle, distinct states which have the same filling fraction.
In particular, new states are generated by transferring
{\it intra}-plane flux-particle attachment, determined by $s_1$ and
$s_2$, to {\it inter}-plane flux attachment, determined by $n$.
Since the inter-plane attached fluxes are negative in these states,
the particles of one plane see the particles of the other plane as
if they were holes. Thus, in these states, there is an effective
attractive force between particles on different planes. Therefore
the wave functions, instead of having zeros of higher order
(which represent repulsion), should have a larger weight when
particles  from different planes are closer to each other
(in the sense of their coordinates projected on the $xy$ plane).
Asymptotically, these wave functions appear to have ``poles"
in the interlayer coordinates. One could think of this state as a
condensate of  inter-layer bound states. Such states should become
preferable as the  interlayer coupling constant becomes large.

\subsection{Spectrum of bilayer FQH states}
\label{subsec:colle}

Following the same steps as in Section \ref{subsec:resfun}, we can
integrate out the gaussian fluctuations of the statistical gauge
fields ${\tilde a}_{\mu}^{\alpha}$, and obtain the effective action for the
electromagnetic fluctuations ${\tilde A}_{\mu}^{\alpha}$
%$S_{\rm eff}^{\rm em}({\tilde A}_{\mu}^{\alpha})$
\begin{equation}
{\cal S}_{\rm eff}^{\rm em} ({\tilde A}_{\mu}^{\alpha}) = {1\over 2}
                  \int d^{3}x \int d^{3}y {\tilde A}_{\mu}^{\alpha}(x)
                  K^{\mu \nu}_{\alpha\beta}(x,y)
                   {\tilde A}_{\nu}^{\beta} (y)
\label{eq:ems}
\end{equation}
Here $K^{\mu\nu}_{\alpha\beta}$ is the electromagnetic polarization
tensor, which measures the linear response of the system to a weak
electromagnetic perturbation. As for the single-layer systems,
we can use this effective action to calculate the full electromagnetic response
functions at the gaussian level.

In this Section we discuss the spectrum of excitations for two
different states, the $(m,m,n)$ and the $(m,m,m)$ states. The
derivation of these results can be found in reference \cite{lopez5}.

We first discuss the case of the $(m,m,n)$ states.
For simplicity, we study the collective modes for the state
$(3,3,1)$. All the other states can be studied by straightforward
application of the same methods \cite{lopez5}.

In this case the total filling fraction is $\nu = {1\over 2}$,
being ${\nu}_{1} = {\nu}_{2} = {1\over 4}$ . The effective
cyclotron frequencies and magnetic fields are
${\omega}_{\rm eff}^{1}={\omega}_{\rm eff}^{2}=
{\bar \omega}= {{\omega}_{c} \over 4}$ and
${B}_{\rm eff}^{1}={B}_{\rm eff}^{2}={\bar B}= {B \over 4}$.

We find that there is a family of collective modes whose zero-momentum
gap is $k {\bar \omega}$, where $k$ is an integer number different from
$1$. At mean field level, there are two modes for each
integer multiple of ${\bar \omega}$. After including the gaussian
fluctuations
we find
that there are no modes with a  zero momentum gap at ${\bar \omega}$.
One of them has been ``pushed up" to the cyclotron frequency and the
other up to $2 {\bar \omega}$ (at ${\vec Q}=0$). Therefore, at these
multiples of  $ {\bar \omega}$ there are three degenerated modes
for ${\vec Q}=0$.
For ${\vec Q} \not=0$, the degeneracy is lifted and these three modes have
different dispersion curves.

At $2 {\bar \omega}$ we find that there are two modes with residue
${\vec Q}^2$, and one with residue ${\vec Q}^4$. The former are
\begin{equation}
\omega_{\pm} ({\vec Q})= \Big [ (2 {\bar \omega})^2 +
                         ({{\vec Q}^2\over 2 {B_{\rm eff}}})^{1\over 2}\;
                         {\bar \omega}^2 \; {\alpha }_{\pm}
             \Big ] ^{1\over2}
\label{eq:uno}
\end{equation}
where
\begin{equation}
{\alpha}_{\pm} ={3M\over 2\pi} ({v}_{11}-{v}_{12}) \pm
            {\sqrt {({3M\over 2\pi})^{2}({v}_{11}-{v}_{12})^2 +16}}
\label{eq:unosub}
\end{equation}
Here $v_{\alpha\beta}$ are the zeroth order coefficients of the
Fourier transform of the interparticle pair potential for short range
interactions. For Coulomb interactions $ v_{11}= {q^2 \over \epsilon}$ and
$ v_{12}= {q^2 \over \epsilon}{e^{-|{\vec Q}|d}}\approx {q^2 \over \epsilon}$
if $|{\vec Q|}d \ll 1$, therefore ${\alpha }_{\pm}= \pm 4$ in this limit.

The residues in $K_{00}^{\alpha\beta}$ corresponding to these poles are
\begin{equation}
Res(K_{00}^{\alpha\beta},\omega _{\pm}({\vec Q})) = -{\vec Q}^2 \;
                                   {\omega}_{c} {\nu \over 8\pi}\;
            \left( \begin{array}{cc} 1& \;  -1\\ -1& \; 1
                   \end{array} \; \right)
                     ( 1+ {16\over {\alpha}_{\pm}^{2}} )^{-1}
\label{eq:resuno}
\end{equation}
It is clear from the form of these residues, that
these excitations are {\it out of phase} modes, because they only couple
the ``out of phase" density ({\it i.~e.\/},
they couple ${\rho }^{-}$ with itself).

The other mode at $2 {\bar \omega}$ is
\begin{equation}
\omega_{0} ({\vec Q})= \Big [ (2 {\bar \omega})^2 -
                         6 ({{\vec Q}^2\over 2 {B_{\rm eff}}})\;
                         {\bar \omega}^2
             \Big ] ^{1\over2}
\label{eq:dos}
\end{equation}
and its residue is proportional to ${\vec Q}^4 \left( \begin{array}{cc} 1& \;
  1\\ 1& \; 1   \end{array} \; \right) $. Thus, this is an {\it in phase}
mode since it only couples ${\rho }^{+}$ with itself.

There are three modes whose zero momentum frequency is the cyclotron
frequency. The only mode with residue ${\vec Q}^2$ has energy
\begin{equation}
\omega ({\vec Q})= \Big [ { \omega}_{c}^2 +
                         ( 2+ {M( v_{11} + v_{12}) \over \pi})
                        3  ({{\vec Q}^2\over 2 {B_{\rm eff}}})\;
                         {\bar \omega}^2   \Big ] ^{1\over2}
\label{eq:tres}
\end{equation}
and residue
\begin{equation}
Res(K_{00},\omega ({\vec Q})) = -{\vec Q}^2 \;
                                   { \omega}_{c} {\nu \over 8\pi}\;
            \left( \begin{array}{cc} 1& \;  1\\ 1& \; 1
                   \end{array}\; \right) ,
\label{eq:restres}
\end{equation}
As in the case of single layers, except for the lowest branch of
collective modes and the cyclotron mode at ${\vec Q}=0$, all the other
modes become damped due to processes induced by
the non-quadratic interactions among the collective modes.

We now discuss the collective modes of the $(m,m,m)$ states.
In this case the total filling fraction is $\nu = {1\over m}$,
being ${\nu}_{1} = {\nu}_{2} = {1\over 2m}$ . The effective
cyclotron frequencies and magnetic fields are
$\; {\omega}_{\rm eff}^{1}={\omega}_{\rm eff}^{2} \equiv
{\bar \omega}= {{\omega}_{c} \over 2m}$ and
${B}_{\rm eff}^{1}={B}_{\rm eff}^{2} \equiv {B _{\rm eff}} =
{B \over 2m}$.

We find again a family of collective modes whose zero-momentum
gap is $k {\bar \omega}$, where $k$ is an integer number different from
$1$. At mean field level, there are two modes for each
integer multiple of ${\bar \omega}$. After including the gaussian
fluctuations
we find
that there are no modes with a  zero momentum gap at ${\bar \omega}$.
One of them has been ``pushed up" to the cyclotron frequency.
Therefore, at
$ {\omega}_{c}$ there are three degenerate modes for ${\vec Q}=0$.
For ${\vec Q} \not=0$, the degeneracy is lifted and these three modes have
different dispersion curves.
The other mode at ${\bar \omega}$
has been ``pulled down"  to zero frequency at ${\vec Q}=0$, {\it i.~e.\/}, it
has become
a gapless mode.

We will distinguish between the cases $m=1$ and $m \not= 1$.

\noindent a) Case $m=1$

The gapless mode is
\begin{equation}
\omega ({\vec Q})= {v_{s}} |{\vec Q}|
\label{eq:veinte}
\end{equation}
where
\begin{equation}
{v_{s}}^{2}=\Big [ 1+ {M\over 2\pi}(v_{11} - v_{12})\Big ]
{ {\omega}_{c} \nu \over 2M}
\label{eq:vel}
\end{equation}
where $v_{\alpha\beta}$ are the zeroth order coefficient of the
Fourier transform of the interparticle pair potential for short range
interactions.
For Coulomb interactions $ v_{11}({\vec Q})= {q^2 \over \epsilon}$ and
$ v_{12}({\vec Q}) = {q^2 \over \epsilon}{e^{-|{\vec Q}|d}}\approx
{q^2 \over \epsilon}$
if $|{\vec Q}|d \ll 1$. Therefore, $(v_{11} - v_{12})=0$ for Coulomb
interactions (in the limit $|{\vec Q}|d \ll 1$).

The residue in $K_{00}^{\alpha\beta}$ corresponding to this pole is
\begin{equation}
Res(K_{00},\omega ({\vec Q})) = -{\vec Q}^2 \;
                                   {\omega}_{c} {\nu \over 8\pi}\;
            \left( \begin{array}{cc} \;1&   -1\\ -1& \; 1
                   \end{array} \; \right)
\label{eq:resveinte}
\end{equation}
Therefore, this is an {\it out of phase} mode.

At ${\omega }_{c}=2 {\bar \omega}$ we find that there are two
({\it in phase}) modes
with residue ${\vec Q}^2$
\begin{equation}
\omega_{\pm} ({\vec Q})= \Big [  {\omega}_{c}^2 +
                         ({{\vec Q}^2\over 2 {B_{\rm eff}}})^{1\over 2}\;
                         {\bar \omega}^2 \; {\alpha }_{\pm}
             \Big ] ^{1\over2}
\label{eq:veintiuno}
\end{equation}
where
\begin{equation}
{\alpha}_{\pm} = {M\over 2\pi}({v}_{11}+{v}_{12}) \pm
            {\sqrt {({M\over 2\pi})^{2}({v}_{11}+{v}_{12})^2 +16}}
\label{eq:veintiunosub}
\end{equation}
 For Coulomb interactions $ v_{11} +v_{12}=  2 {q^2 \over \epsilon}$ if
$|{\vec Q}|d \ll 1$, therefore this term is higher order in $\vec Q$
and it should be neglected, {\it i.~e.\/}, ${\alpha }_{\pm}=4$.

The residues in $K_{00}^{\alpha\beta}$ corresponding to these poles are
\begin{equation}
Res(K_{00},\omega _{\pm}({\vec Q})) = -{\vec Q}^2 \;
                                   {\omega}_{c} {\nu \over 8\pi}\;
            \left( \begin{array}{cc} 1& \;  1\\ 1& \; 1
                   \end{array} \; \right)
                     ( 1+ {16\over {\alpha}_{\pm}^{2}} )^{-1}
\label{eq:resveintiuno}
\end{equation}
therefore these are {\it in phase} modes.
The other mode at $\omega _{c}$ is
\begin{equation}
\omega_{0} ({\vec Q})= \Big [ {\omega}_{c}^2 - 2
                       ({{\vec Q}^2\over 2 {B_{\rm eff}}})\;
                         {\bar \omega}^2      \Big ] ^{1\over2}
\label{eq:veintidos}
\end{equation}
and its residue is proportional to ${\vec Q}^4
\left( \begin{array}{cc} \; 1&  - 1\\ -1& \; 1
                   \end{array} \; \right)$ ({\it out of phase} mode).

\noindent b) Case $m \not= 1$

The gapless mode is an {\it out of phase} mode with the same form as
for $m=1$ (Eq.~\ref{eq:veinte})
and with the same residue (Eq.~\ref{eq:resveinte}).

At ${\omega }_{c}=2m {\bar \omega}$ we find that there is only one
{\it in phase} mode with residue ${\vec Q}^2$
\begin{equation}
\omega ({\vec Q})= \Big [  {\omega}_{c}^2 +
              \left( {{2m-1}\over {m-1}}+ {M\over 2\pi}
(v_{11} + v_{12})\right)
                   2m  ({{\vec Q}^2\over 2 {B_{\rm eff}}})\;
                         {\bar \omega}^2               \Big ] ^{1\over2}
\label{eq:treintayuno}
\end{equation}

The residue in $K_{00}^{\alpha\beta}$ corresponding to this pole is
\begin{equation}
Res(K_{00},\omega ({\vec Q})) = -{\vec Q}^2 \;
                                   {\omega}_{c} {\nu \over 8\pi}\;
            \left( \begin{array}{cc} 1& \;  1\\ 1& \; 1
                   \end{array} \; \right)
\label{eq:restreintayuno}
\end{equation}
The modes with zero momentum frequency $\omega = k {\bar \omega}$ for
$k \geq 3$  have been calculated in reference \cite{lopez5}.

All the considerations about the validity of this spectrum of collective
excitations beyond the semiclassical approximation that we discussed
in previous sections are of course valid in this case as well.

\subsection{Effective Low Energy Action, Hall Conductance and
Quantum Numbers}
\label{subsec:Hall}

We will derive now the effective action for the low energy degrees of
freedom of the bilayer system. Using the same line of argument of
Section \ref{subsec:hallsingle}, we keep only the leading terms in a
gradient expansion and find
\begin{eqnarray}
S_{\rm eff} ({\tilde a}_{\mu}^{\alpha},{\tilde A}_{\mu}^{\alpha})
 && \approx {{\bar \kappa}^{\alpha \beta} \over 2}
                       \int d^3x \; {\tilde a}_{\mu}^{\alpha}
\;{\epsilon _{\mu\nu\lambda}} {\partial^\lambda}
                       \; {\tilde a}_{\nu}^{\beta}
            -{{\kappa}^{\alpha \beta} \over 2}
                       \int d^3x \;
                       {\tilde a}_{\mu}^{\alpha}\;
{\epsilon _{\mu\nu\lambda}} {\partial^\lambda}
                       \; {\tilde A}_{\nu}^{\beta}
\nonumber \\
        && -{{\kappa}^{\alpha \beta} \over 2}
                       \int d^3x \;
                       {\tilde A}_{\mu}^{\alpha}\;
{\epsilon _{\mu\nu\lambda}}
                       {\partial^\lambda}\; {\tilde a}_{\nu}^{\beta}
            +{{\kappa}^{\alpha \beta} \over 2}
                       \int d^3x \;
                       {\tilde A}_{\mu}^{\alpha}\;
{\epsilon _{\mu\nu\lambda}}
                       {\partial^\lambda}\; {\tilde A}_{\nu}^{\beta}
\nonumber \\
        &&
\label{eq:shal}
\end{eqnarray}
where we have set ${\bar \kappa}^{\alpha \beta}= {{p_{\alpha}}\over 2\pi}
{\delta}^{\alpha \beta} + {\kappa}^{\alpha \beta}$.

The effective action of Eq.~\ref{eq:shal} is very similar to the
effective action of Wen and Zee's Matrix classification of Abelian
quantum Hall states \cite{wenzee-matrix}. It differs from it in two
respects: in  Eq.~\ref{eq:shal} there is an induced Chern-Simons term for
the background electromagnetic fields, and the coefficients are not
quite the same as in reference \cite{wenzee-matrix}. However, as we
will see below, Eq.~\ref{eq:shal} leads to the correct value of
the Hall conductance and of the fractional charge and
statistics of the quasiparticles. Thus, they describe the same physics.

The next step is to integrate the statistical gauge fields to obtain the
effective action for the electromagnetic field. In particular,
we will need
to compute the inverse of the matrix ${\bar \kappa}^{\alpha \beta}$. This
inverse  only exists if
$\Delta = [({1\over p_{1}} + 2 s_{1})({1\over p_{2}} + 2 s_{2}) - n^{2}]
\not= 0$. Therefore we must consider two cases, the one in which
$\Delta \not= 0$ and the one in which $\Delta =0$.

\noindent Case $\Delta \not= 0$

Upon integrating over the statistical gauge fields in Eq.~\ref{eq:shal},
the effective action for the electromagnetic field results
\begin{equation}
S_{\rm eff}^{\rm em} ({\tilde A}_{\mu}^{\alpha})  \approx
{{\kappa}_{\rm eff}^{\alpha \gamma} \over 2}
                       \int d^3x
                       {\tilde A}_{\mu}^{\alpha} \;
                       {\epsilon _{\mu\nu\lambda}}
                       { \partial^{\lambda}}\; {\tilde A}_{\nu}^{\beta}
\label{eq:sel}
\end{equation}
where the effective quantum Hall {\it conductance matrix}
${\kappa}_{\rm eff}^{\alpha\beta}$
\begin{equation}
{\kappa}_{\rm eff}^{\alpha\beta}=
{\kappa}^{\alpha \gamma} [{\delta}^{\gamma \beta } -
({\bar \kappa}^{-1})^{\gamma\delta} {\kappa}^{\delta \beta}]
\label{eq:Keff}
\end{equation}
is given by
\begin{eqnarray}
 {\kappa}_{\rm eff}^{\alpha \beta}
={1 \over 2\pi [({1\over p_{1}} + 2 s_{1})
({1\over p_{2}} + 2 s_{2}) - n^{2}]}
\left( \begin{array}{cc}
({1\over p_{2}} + 2 s_{2})  & -n \\
-n & ({1\over p_{1}} + 2 s_{1})
\end{array}    \right)
\label{eq:sig}
\end{eqnarray}
The matrix ${\kappa}_{\rm eff}^{\alpha\beta}$ gives the
intralayer and translayer conductances of the bilayer system. The
translayer conductances characterize the Coulomb drag effects in
bilayers.

In particular, let us consider the case in which both layers are coupled
to the same electromagnetic field. Then ${\tilde A}_{\mu}^1 =
{\tilde A}_{\mu}^2={\tilde A}_{\mu}$, and the coefficient in the
effective action results
\begin{equation}
  {\kappa}_{\rm eff}\equiv
\sum_{\alpha \beta}{\kappa}_{\rm eff}^{\alpha \beta}
={1 \over 2\pi} { {2n- ({1\over p_{1}} + 2 s_{1})-
({1\over p_{2}} + 2 s_{2})}
\over  { n^{2}-({1\over p_{1}} + 2 s_{1})({1\over p_{2}} + 2 s_{2}) }}
={\nu \over 2\pi}
\label{eq:sigma23}
\end{equation}
Following the same arguments as in Section \ref{subsec:hallsingle} we
find that the electromagnetic current induced in the system
is $J_{\mu}={ {{\kappa}_{\rm eff}}\over 2} \epsilon_{\mu\nu \lambda}
{\tilde F}^{\nu\lambda}$.
Thus, if a weak external electric field ${\tilde E}_j$ is applied, the
induced current is
$J_k=  {\kappa}_{\rm eff} \epsilon _{lk} {\tilde E}_l $.
Therefore the coefficient $ {\kappa}_{\rm eff}$ is the {\it
actual} Hall conductance of the system.
\begin{equation}
 \sigma _{xy}\equiv  {\kappa}_{\rm eff} ={\nu\over 2 \pi}
\end{equation}
which is a {\it fractional} multiple of ${e^2\over h}$ (in units in
which $e=\hbar=1$). Thus, the uniform states exhibit a Fractional
Quantum Hall effect with the correct value of the Hall conductance.

The effective action of Eq.~\ref{eq:shal} also
determines the quantum numbers
({\it i.~e.\/}, charge and statistics)
of the vortices (topological excitations) supported by this FQH state.
These quantum numbers were calculated in reference \cite{lopez5} and we
will not give an explicit derivation here. The charge of the excitations
was found to be given by
\begin{equation}
q_{\rm eff}^{\alpha} = { {\nu}^{\alpha} \over p^{\alpha} }
\label{eq:effq}
\end{equation}
In particular, for the ($m,m,n$) states, the effective charge in both
layers is the same and is given by $q_{\rm eff}= {1\over {n+m}}$.

On the other hand, the statistics of the excitations on the same layer is
\begin{equation}
\delta_{\alpha}= \pi \big( 1+ {(p_{\beta}+
                       { 2s_{\alpha}\over {4s_{1}s_{2}- n^{2}}   } )\over
{ (p_{1}+ { 2s_{2}\over {4s_{1}s_{2}- n^{2}}   } )
(p_{2}+ { 2s_{1}\over {4s_{1}s_{2}- n^{2}}   } )
- {n^{2}\over ({4s_{1}s_{2}- n^{2}})^{2}  }}} \big)
\label{eq:sss1}
\end{equation}
where if $\alpha=1$, then $\beta=2$ and viceversa.
In particular, for the ($m,m,n$) states, the statistics is $\delta = -{m
\over {(m^2 -n^2)}}$, independent of the layer.
The {\it relative} statistics ({\it i.~e.\/}, for excitations on
different layers)
$\delta _{12}$ is given by
\begin{equation}
\delta_{12}= \pi  {{n\over {4s_{1}s_{2}- n^{2}}} \over
{ (p_{1}+ { 2s_{2}\over {4s_{1}s_{2}- n^{2}}   } )
(p_{2}+ { 2s_{1}\over {4s_{1}s_{2}- n^{2}}   } )
- {n^{2}\over ({4s_{1}s_{2}- n^{2}})^{2}  }}}
\label{eq:sss12}
\end{equation}
In particular, for the ($m,m,n$) states, the {\it relative} statistics
results $\delta ={n \over {(m^2 -n^2)}}$.

It is worth to discuss these results in the context of Haldanes's
topological stability criterion \cite{haldane-stability}.
After some (tedious) algebra it is straightforward to show that
Haldanes's topological stability is
equivalent to the requirement that  the denominators in Eq.~\ref{eq:sss1}
be non-zero. For the special case of the ($m,m,n$) states, this holds if
$m \neq n$. We will see next that this requirement fails if there exists
one branch of collective excitations with a gapless mode, such as in the
($m,m,m$) states.

\noindent Case $\Delta = 0$

It can be shown that when ${\bar \kappa}^{\alpha\beta}$ is not invertible,
{\it i.~e.\/}, it has a zero eigenvalue, the corresponding linear
combination of the gauge fields become massless. In other words,
the {\it in phase} gauge field ${\tilde a}_{\mu}^{+}=
{\tilde a}_{\mu}^{1}+{\tilde a}_{\mu}^{2}$ has a finite gap which couples
to the electromagnetic field  ${\tilde A}_{\mu}$, while the {\it out of phase}
gauge field ${\tilde a}_{\mu}^{-}=
{\tilde a}_{\mu}^{1}-{\tilde a}_{\mu}^{2}$ is gapless.

We will study in particular the case of the ($m,m,m$) states which satisfy the
condition $\Delta=0$.
For these states
\begin{eqnarray}
 {\bar \kappa}={-m  \over 2\pi (1-2m)}
\left( \begin{array}{cc}
1  & 1 \\
1 & 1
\end{array}    \right)
\label{eq:kapa}
\end{eqnarray}
which is clearly non invertible.

We can write the effective action for the fluctuations of the Chern-Simons
gauge fields (Eq.~\ref{eq:shal}) in the basis defined by
${\tilde a}_{\mu}^{\pm} ={\tilde a}_{\mu}^{1}\pm {\tilde a}_{\mu}^{2}$ .
\begin{eqnarray}
S_{\rm eff} ({\tilde a}_{\mu}^{\alpha},{\tilde A}_{\mu}^{\alpha})
  \approx &&\!\!\!\!\!\!\!\!\!-{1\over 2} \int d^{3}x \;
 {m\over {2\pi (1-2m)}} {\tilde a}_{\mu}^{+} \;{\epsilon _{\mu\nu\lambda}}
                       { \partial^{\lambda}}\; {\tilde a}_{\nu}^{+}
\nonumber \\
         &&\!\!\!\!\!\!\!\!\!\!\!\!\!\!\!\!\!\!\!\!\!\!
-{1\over 2} \int d^{3}x \;
 {1\over {4\pi (1-2m)}}(-{\tilde a}_{\mu}^{+}{\epsilon _{\mu\nu\lambda}}
                       { \partial^{\lambda}}\; {\tilde A}_{\nu}^{+}
                         + (2m-1)
                       {\tilde a}_{\mu}^{-}{\epsilon _{\mu\nu\lambda}}
                       { \partial^{\lambda}}\; {\tilde A}_{\nu}^{-})
\nonumber \\
         &&\!\!\!\!\!\!\!\!\!\!\!\!\!\!\!\!\!\!\!\!\!\!
-{1\over 2} \int d^{3}x \;
 {1\over {4\pi (1-2m)}}(-{\tilde A}_{\mu}^{+}{\epsilon _{\mu\nu\lambda}}
                       { \partial^{\lambda}}\;  {\tilde a}_{\nu}^{+}
                         + (2m-1)
                       {\tilde A}_{\mu}^{-}{\epsilon _{\mu\nu\lambda}}
                       { \partial^{\lambda}}\; {\tilde a}_{\nu}^{-})
\nonumber \\
        &&\!\!\!\!\!\!\!\!\!\!\!\!\!\!\!\!\!\!\!\!\!\!
+{1\over 2}\int d^{3}x \;
 {1\over {4\pi (1-2m)}}(-{\tilde A}_{\mu}^{+}{\epsilon _{\mu\nu\lambda}}
                       { \partial^{\lambda}}\; {\tilde A}_{\nu}^{+}
                         + (2m-1)
                       {\tilde A}_{\mu}^{-}{\epsilon _{\mu\nu\lambda}}
                       { \partial^{\lambda}}\; {\tilde A}_{\nu}^{-})
\nonumber\\
\label{eq:shal2}
\end{eqnarray}
The gauge field $ {\tilde a}_{\nu}^{-}$ appears as a Lagrange multiplier
in
this action. The integration over it states that the current
${\epsilon _{\mu\nu\lambda}} { \partial^{\lambda}}\; {\tilde
A}_{\nu}^{-}$ vanishes.
This is trivially valid if the electromagnetic field is the same for both
layers, because $ {\tilde A}_{\nu}^{-}=0$.

The integration over $ {\tilde a}_{\nu}^{+}$ gives the effective action
for the
field $ {\tilde A}_{\nu}^{+}$. If we consider the case in which
${\tilde A}_{\mu}^1 ={\tilde A}_{\mu}^2={\tilde A}_{\mu}$, the result is
\begin{equation}
S_{\rm eff}^{\rm em} ({\tilde A}_{\mu}) =
{1\over 2\pi m}    \int d^3x
                       {\tilde A}_{\mu}       {\epsilon _{\mu\nu\lambda}}
                       { \partial^{\lambda}}{\tilde A}_{\nu}
\label{eq:sel1}
\end{equation}
Following the same steps as in the previous case, the Hall conductance
results
$\sigma _{xy}= {1\over 2\pi m}={\nu \over 2\pi}$ which is the correct
value
for the ($m,m,m$) states.

The existence of a gapless branch in the spectrum of collective
excitations has important consequences for the spectrum of topological
excitations {\it i.~e.\/}, quasiparticles and quasiholes.
It was shown in reference \cite{lopez5} that the gapless collective
excitations mediate an effective confining gauge force between
topological excitations which carry the quantum numbers of the 
out-of-phase (or inter-layer) mode. The result is that the only 
surviving excitations are bound states which are neutral with respect 
to the inter-layer gauge group but carry charge $1/m$ and fractional 
statistics $\pi/m$ (measured from fermions). These states have a 
finite energy gap.
For the particular case of the state $(1,1,1)$, these bound states are
{\it bosons} of charge $1$ and finite gap. This gapped bound state has
the same quantum numbers as the {\it skyrmion} excitation of the $(1,1,1)$
discussed by K.~Yang {\it et.~al.\/}, who view the $(1,1,1)$ state as a
quantum ferromagnet \cite{indiana}. S.~L.~Sondhi and collaborators had
predicted earlier that a partially polarized $\nu=1$ QH state should
have skyrmion states \cite{skkr}. Our results indicate that all of
the  $(m,m,m)$ states should have skyrmion excitations but, in
general, they have fractional charge $1/m$ and fractional statistics 
$\pi/m$. Hence, the bilayer state $(3,3,3)$, which occurs at $\nu=1/3$
($1/6$ on each layer), has skyrmions of charge $1/3$ and statistics $\pi/3$.

The reduction of the number of topological excitations for the case
$m=n$ illustrates, in a physically
transparent way, the meaning of Haldane's criterion for topological
stability. Indeed, Haldane's criterion predicts that the $(m,m,m)$ states
are not topologically stable. This of course does not mean that the state
itself is unstable or unphysical. Rather, it means that if one or more
branches of collective excitations become gapless, the quasiparticles
which couple to them are not present in the spectrum and the corresponding
quantum numbers are not allowed.

\subsection{Spin Singlet States and $SU(2)$ Symmetry}
\label{subsec:su2}

In this Section we discuss the application of Chern-Simons methods to a
2DEG with an $SU(2)$ symmetry. This symmetry may be exact or may be
broken by explicit terms in the Hamiltonian.
In reference \cite{su2}
a non-abelian Chern-Simons approach was used to construct a theory of
the spin singlet $(m+1 , m+1, m)$ Halperin states. Here, we present
a generalization of the approach of reference \cite{su2} which will yield
the $SU(2)$ hierarchies. A related, and completely equivalent
approach was developed independently
by Frohlich, Kerler, and Marchetti \cite{fkm}.

In the approach of Balatsky and Fradkin \cite{su2} (BF), instead of the
two-component {\it abelian} Chern-Simons theory that we use in this
Chapter, a {\it non-abelian} Chern-Simons gauge field is introduced.
The advantage of the BF approach is that the $SU(2)$ invariance is
manifest and it is not the consequence of a subtle dynamical mechanism.
The disadvantage of the BF approach is that the non-abelian Chern-Simons
theory is substantially more sophisticated and technically more
demanding than the abelian theory used before.
In the BF approach, the {\it
electron} is viewed as a composite object which is made of a particle
that carries the charge (the {\it holon}) and another particle that
carries the spin (the {\it spinon}). This {\it arbitrary} separation
gives rise to the existence of an abelian gauge symmetry (called {\it
RVB} by BF). The requirement of gauge invariance forces the holons and
spinons to be glued together in bound states, the electrons. In the BF
approach, the need of a non-abelian gauge field is a consequence
of the assignment of fractional (semion) statistics to both
holons and spinons. In this way,
$SU(2)$ fluxes are attached to a set of charge neutral,
spin-${\frac{1}{2}}$,
fermions, which become the spinons. The $SU(2)$ symmetry only admits
Bose, Fermi or semionic statistics. The holons, instead, are represented
by fermions attached to $U(1)$ fluxes. The $U(1)$ and $SU(2)$
Chern-Simons coupling constants must be chosen in such a way that the
holons and spinons are semions. Within this approach, the FQHE of the
spin-singlet states is the FQHE of the semions. The spin structure just
sits on top of the FQHE.

There is another, and more obvious, way to attach fluxes to particles
while keeping the $SU(2)$ spin symmetry untouched.
Belkhir and Jain \cite{jainspin} proposed a spin singlet wave
function for the
state with filling fraction $\nu={\frac{1}{2}}$, based on a composite
fermion picture. In their construction, they attach the {\it same}
number of pairs of flux quanta to {\it both} up and down spins.
These approaches are
represented in
the $U(1) \otimes U(1)$ theory of the preceeding sections
by demanding that both up and down electrons see that same
flux at all times. This means to choose a
Chern-Simons matrix ${\kappa}^{\alpha \beta}$ which is proportional to the
identity, {\it i.~e.\/}, $s_1=s_2$ and $n=0$. It is easy to use this
approach to get the $SU(2)$ limit of the Jain states, but with arbitrary
polarization. However, it is not a very efficient approach to get all of
the spin singlet states. The approach of BF deals with this states more
directly.
Alternatively, it is possible to attach flux to the charge degrees of
freedom, leaving spin untouched. This
construction, but in the bosonic Chern-Simons language, was used
by D.~H.~Lee and C.~Kane \cite{leekane} and subsequently applied by
Sondhi, Karlhede, Kivelson and Rezayi \cite{skkr} in their theory of
skyrmion states in polarized FQHE states. This approach does not
describe the spin singlet states.
In what follows, we follow the BF approach.

More concretely, we introduce a holon field $\phi$ and a
spinon field $\chi_\alpha$ ($\alpha=\uparrow,\downarrow$) and represent
the {\it
electron} field operator $\psi_\alpha$ as $\psi_\alpha(x)=\phi(x) \;
\chi_\alpha(x)$. BF showed that, for an $SU(2)$ invariant system, the
system defined by the action of Eq.~\ref{eq:ese1} is equivalent to the
following theory of (interacting) spinons and holons
\begin{equation}
{\cal S}= {\cal S}_{\rm charge}+ {\cal S}_{\rm spin} + {\cal S}_{\rm
interaction}
\end{equation}
where the action for the charge degrees of freedom is
\begin{eqnarray}
{\cal S}_{\rm charge}&=&\int d^3x\;\left( \phi^{\dagger}(x)\; (
iD_0^c+\mu)
\; \phi(x)
+{\frac{1}{2M}} \phi^{\dagger}(x)\; {\vec D_c}^2 \phi(x)
\right)
\nonumber\\
&+&\int d^3x\;{\frac{\theta}{2}} \epsilon_{\mu \nu \lambda}
a^\mu(x) \partial^\nu a^\lambda(x)
\label{eq:scharge}
\end{eqnarray}
while the action for spin is
\begin{eqnarray}
{\cal S}_{\rm spin}=&\int& d^3x\;\left( \chi_\alpha^{\dagger}(x)\;
iD_0^s
\; \chi_\alpha(x)+{\frac{1}{2M}} \chi_\alpha^{\dagger}(x)\; {\vec D_s}^2
\chi_\alpha(x)
\right)\nonumber\\
-&\int& d^3x \; {\frac{k}{4 \pi}} \epsilon_{\mu \nu
\lambda}
{\rm tr}( b^\mu(x) \partial^\nu b^\lambda(x)+{\frac {2}{3}} b^\mu
(x) b^\nu (x) b^\lambda (x))
\label{eq:sspin}
\end{eqnarray}
and an (instantaneous) pair interaction term
\begin{equation}
{\cal S}_{\rm interaction}=-\int d^3x \; \int d^3x' \; {\frac{1}{2}}
(\rho(x)-{\bar \rho}) V(x-x') (\rho(x')-{\bar \rho})
\label{eq:sint}
\end{equation}
where $\rho(x)$ is the charge density
\begin{equation}
\rho(x)=\psi_\alpha^\dagger(x)\psi_\alpha(x)\equiv \phi^\dagger(x)
\phi(x)
\label{eq:charge}
\end{equation}
(see below).
In Eq.~\ref{eq:scharge}, $a_\mu$ is the statistical vector
potential which turns the holons into semions. This condition requires
that the $U(1)$ Chern-Simons coupling constant $\theta$ be restricted
to the values
\begin{equation}
{\frac{1}{\theta}}=-{\frac{2\pi}{m}}+2\pi 2s
\label{eq:thetasemion}
\end{equation}
where semion statistics requires that $m=\pm 2$ and $s$ is an arbitrary
integer. Likewise, in Eq.~\ref{eq:sspin} $b_\mu$ is the $SU(2)$
non-abelian statistical gauge field which takes values on the $SU(2)$
algebra. Hence, we can expand the field  in the form
$b_\mu(x)=b_\mu^a(x)\; \tau^a$, where
$\tau^a$ ($a=1,2,3$) are the three generators of $SU(2)$ in the spinor
representation , {\it i.~e.\/}, the set of $2 \times 2$ Pauli matrices.
The $SU(2)$ Chern-Simons coupling constant $k$, the {\it level}
of the Chern-Simons theory is , for our system, equal to $k=\pm 1$
\cite{su2}.
Hence, we have a (``trivial") level one Chern-Simons theory. All the
representations of a level one Chern-Simons theory are known to be
abelian and to correspond to abelian fractional statistics of fermions,
bosons or semions \cite{witten}. The only ambiguity left is the sign of
$k$ which is the chirality (or handedness) of the semion. There is a
similar sign ambiguity in the coupling constant $\theta$ in
Eq.~\ref{eq:thetasemion}. Different choices of these signs lead to
different statistics for the bound states of holons and spinons. The
requirement that the bound state be an electron, which is a fermion,
leads to the condition ${\rm sign}(k)={\rm sign}(m)$.

The $U(1)$ and $SU(2)$ charge and spin covariant derivatives are
\begin{eqnarray}
D_\mu^c&=&\partial_\mu-i(A_\mu+a_\mu+c_\mu)\nonumber\\
D_\mu^s&=&\;I\;\partial_\mu-i(b_\mu-I\;c_\mu)
\label{eq:covder}
\end{eqnarray}
where $A_\mu$ is the external electromagnetic field and $I$ is the $2
\times 2$ identity matrix. The gauge field $c_\mu$ is the
$U(1)$ ``{\em RVB}" gauge field which glues spins and charges together.
The covariant derivatives have been chosen in such a way that
holons and spinons have opposite charge with respect to $c_\mu$. Hence,
the strong fluctuations of this field binds holons and spinons into
states which are locally singlets under the ``RVB" gauge
transformations, {\it i.~e.\/}, electrons. In fact, all the gauge field
$c_\mu$ does is to enforce the constraint that the 3-current
of the holons equals the 3-current of the spinons, as an operator
statement in the physical Hilbert space. This is seen clearly from the
equation of motion generated by $c_\mu$
\begin{equation}
J^{\rm RVB}(x)=\frac{\delta {\cal S}}{\delta c_0(x)}=0 \;
\Longrightarrow\;\phi^{\dagger}(x)\phi(x)-\sum_{\sigma=\uparrow,
\downarrow}\chi_\sigma^{\dagger}(x)\chi_\sigma(x)=0
\label{eq:jrvb}
\end{equation}
Thus $N_{\rm e}$, the number of charges, and $N_{\uparrow}$ and
$N_{\downarrow}$, the number of up and down spins, must obey the obvious
relation $N_{\rm e}=N_{\uparrow}+N_{\downarrow}$. Also,
Eq.~\ref{eq:jrvb} tells us that the local particle density
operator $\rho(x)$ can be identified with the holon charge density
operator, {\it i.~e.\/}, $\rho(x)= \phi^{\dagger}(x)\phi(x)$.

In the way the theory has been set up, it may appear that this
theory may apply only if $SU(2)$ is an exact symmetry. However, this is
not the case. Let us consider for instance the effects of an
$SU(2)$ symmetry breaking field of the form of a Zeeman term in the action
\begin{equation}
{\cal S}_{\rm Zeeman}=\int d^3x \; {\vec B} \cdot
\psi_\alpha^\dagger(x){\vec \tau}_{\alpha \beta} \psi_\beta(x)
\label{eq:zeeman1}
\end{equation}
Using the definition $\psi_\alpha(x)=\phi(x) \; \chi_\alpha(x)$ and the
constraint Eq.~\ref{eq:jrvb}, we can write the Zeeman term in the
equivalent form
\begin{equation}
{\cal S}_{\rm Zeeman}= \int d^3x \; {\vec B} \cdot
\chi_\alpha^\dagger(x){\vec \tau}_{\alpha \beta} \chi_\beta(x)
\label{eq:zeeman2}
\end{equation}
Hence, the Zeeman operators of Eq.~\ref{eq:zeeman1} and
Eq.~\ref{eq:zeeman2} have exactly the same form. This is an exact
identity in the restricted Hilbert space of states satisfying the
constraint of Eq.~\ref{eq:jrvb}.
Likewise, spin-spin pair interaction exchange terms can be written
exactly in the same way. Clearly, these rules also apply to any term
involving spins even if the symmetry is not respected. Thus,
a bilayer 2DEG is a system of fermions with a layer label which can
be regarded as a ``spin" index, even though the $SU(2)$ symmetry is
actually broken. The interlayer tunneling term can be regarded as
Zeeman term. Similarly, the matrix pair interaction of the electrons in
the bilayer system can be represented, in the $SU(2)$ formulation, in
terms of a
suitable combination of the $SU(2)$ invariant charge pair interaction
and an Ising interaction term.

We will not go into the details of the full non-abelian theory.
Instead, we will use this theory to determine the allowed
fractions for $SU(2)$ invariant states. Thus, we will just consider the AFA
equations
for this theory. There are two sets of AFA equations, one for the charge
sector and one for the spin sector. The AFA equations for the charge
sector are just the AFA equations for a charged interacting semion
liquid with $N_{\rm e}$ particles in an uniform magnetic field with
$N_\phi$ flux quanta, {\it i.~e.\/}, a FQHE of semions. A simple
application of the methods of Section \ref{subsec:CS} yields the constraint
for the charge density operator
\begin{equation}
\frac{\delta {\cal S}}{\delta a_0(x)}=0 \;\Longrightarrow \;
 j_0(x) =-\theta {\cal B}_c (x)
\label{eq:semioncharge}
\end{equation}
where $ j_0(x)$ is the charge density in the
liquid state, {\it i.~e.\/}, ${\bar \rho}$, and ${\cal B}_c(x)$
is the $U(1)$ statistical charge flux. This equation, when
specialized on fluid states, yields
the allowed fractions. Similar considerations for the $SU(2)$ gauge
field $b_\mu^a$ (with $a=1,2,3$) yield a constraint for the local spin
density operator
\begin{equation}
{\frac{\delta {\cal S}}{\delta b_\mu(x)^a}}=0 \; \Longrightarrow \;
 j_0^{a}(x) =-
\psi_\sigma^{\dagger} \tau^a \psi_\sigma={\frac{k}{2
\pi}}{\cal B}_s^{a}
\label{eq:semionspin}
\end{equation}
where ${\cal B}_s^{a}$ is the $SU(2)$ spin flux.

For fluid states, we generalize the Average Field
Approximation and replace these exact local operator
identities by translationally invariant averages. This replacement may
be problematic in the case of the $RVB$ gauge field since its only
mission is to enforce {\it exactly} the constraint Eq.~\ref{eq:jrvb}.
At the level of the wave functions for the allowed ground states, this
constraint simply means that every coordinate for a charge degree of
freedom has to coincide with the coordinate of a spin degree of freedom.

Thus, we seek fluid states with $N_e$
particles, $N_\phi$ flux quanta and filling
fraction $\nu=N_e/N_\phi$. We wish to determine the filling
fractions and spin for which the ground state is a fluid.
In addition to the usual AFA equation
for the charge, that will give the allowed fractions, we will now get
conditions for the spin and polarization of the allowed states.

In the charge sector, we have a FQHE of spinless semions. In this case,
the AFA consists of  a system of $N_e$ spinless fermions filling up
effective Landau levels, exactly as in the case of spin
polarized electrons (and anyons!). The same line of argument
that was used to derive the allowed fractions in Section
\ref{subsec:CS}
now tells us that, from Eq.~\ref{eq:semioncharge}, the effective number of
fluxes is
\begin{equation}
{\bar N}_{\phi}=N_{\phi}-{\frac{N_e}{2 \pi \theta}}
\label{eq:fluxsu2}
\end{equation}
and, hence,
the allowed fractions, $\nu^{\pm}$, for the $SU(2)$ fluid states satisfy
\begin{equation}
{\frac{1}{\nu^{\pm}}}=\pm {\frac{1}{p}}-{\frac{1}{2 \pi \theta}}
\label{eq:su2frac}
\end{equation}
where $p$ is a positive integer and the $+$ sign corresponds to a
``particle-like" FQHE  ({\it i.~e.\/}, the effective flux parallel to the
external flux)
while the $-$ sign holds for a ``hole-like" FQHE ({\it i.~e.\/}, the
effective flux anti-parallel to the external flux) . By using the allowed
values of $\theta$ we find that the allowed fractions for the
$SU(2)$ fluid states are of the form
\begin{equation}
\nu^{\pm}(p,s;m)={\frac{mp}{(2sm-1)p+m}} \equiv {\frac{2p}{\pm 2 +
(4s-{\rm sign}(m))p}}
\label{eq:allowed}
\end{equation}
where we have specialized for the case of interest, $m=2 {\rm sign}(m)$.

The hierarchy of FQHE states of Eq.~\ref{eq:allowed} is a
generalization of the states found
by BF, which are obtained by setting $p=+1$ and $m=+2$, {\it i.~e.\/},
$\nu^+(1,s;-2)={\frac{2}{4s+1}}=2,{\frac{2}{5}},{\frac{2}{9}}, \ldots$,
which coincide with the Halperin spin singlet states.
The state with $\nu = {1\over 2}$ is also part of the hierarchy of
Eq.~\ref{eq:allowed}, where it is realized as the state with
$p=+2$ ($-2$) for $m=+2$($-2$) and $s=1$. Numerical studies show that,
for Coulomb-like interactions, this state is not favored and that the
($3,3,1$) state is an accurate representation of the ground state.
Notice that this hierarchy includes a state at filling
fraction $\nu={5\over 2}$. This state is found as the $p=-10$
(``hole-like"), $m=-2$ and $s=0$ member of the $SU(2)$ hierarchy, or as
the $p_1=p_2=-5$, $s_1=s_2=s$ and $n+2s=1$ member of the
$U(1) \otimes U(1)$ hierarchy.

In contrast with the states found in Section \ref{subsec:CS}
using the abelian theory for bilayers,
the states of the $SU(2)$ theory {\it at each filling fraction}, are
arranged in irreducible representations (multiplets) of $SU(2)$.
The spin and polarization of the states is determined by
Eq.~\ref{eq:semionspin}. The only subtlety here is that, since the
components of the total spin do not commute with each other, one can
only determine the total spin and total projection. For a system without
a boundary, the choice of total spin $S$ and total projection along an
arbitrary polarization axis, say $S_z$, are constants of motion which
are invariant under local $SU(2)$ gauge transformations (but, of course,
change under global $SU(2)$ rotations). There are two generic situations
of physical interest: (a) spin {\it singlet} states (or with {\it
microscopic} total spin $S/N_e \approx O(1/N_e)$) and (b) states with
macroscopic spin, $S \approx N_e$, {\it i.~e.\/}, {\it ferromagnetic} states.
The total z-component of the spin polarization
$M={\frac{1}{2}}(N_{\uparrow}-N_{\downarrow})$ obeys
\begin{equation}
2M=N_{\uparrow}-N_{\downarrow}={\frac{k}{2 \pi}} \langle
{\cal B}_3^s \rangle L^2
\label{eq:allowedspin}
\end{equation}
where $L^2$ is the area. It is clear that it is possible to construct
all multiplets with spin $|S| \leq {\frac{N_e}{2}}$. In the
thermodynamic limit, the spin {\it singlet} states have $S=0$ and,
hence, $N_{\uparrow}=N_{\downarrow}=N_e/2$. In contrast, the
ferromagnetic states have, with an appropriate choice of the
quantization axis, a non-vanishing {\it extensive} value of $M$ and,
hence, a non-zero value of ${\cal B}_3^s$.

The wave functions for the spin sector of the spin singlet states have
to be determined from the states of a level one $SU(2)$ Chern-Simons
gauge theory with $N_e$ sources in the fundamental representation. It
was shown by Witten \cite{witten} that these wave functions are
correlation functions of conformal blocks of a conformal field theory in
two Euclidean dimensions, the $SU(2)$ level one Wess-Zumino-Witten
model. This fact was used by Read and Moore \cite{readmoore} and by
Balatsky and Fradkin \cite{su2} to show that, the wave function of the
spin singlet FQHE has a factor which is precisely this conformal block
correlation function. It was also noticed \cite{su2} that this factor
coincides with the Kalmeyer-Laughlin wave function \cite{kl} for a
Spin Liquid.

The states with macroscopic spin have a somewhat different physics. The
existence of a non-zero average field should make a mean field approach
more sound. The spin sector of this mean field theory has
$N_{\uparrow,\downarrow}={\frac {1}{2}}N_e \pm M$ spin up and spin down
spinons each feeling an effective uniform magnetic field of $\pm
{\frac{2 \pi}{k}}M$. Because of the $SU(2)$ invariance,
there is no Zeeman term and only the orbital degrees of freedom see
this spin-dependent external field. It is easy to see that the highest
weight ferromagnetic state with maximal spin is obtained by filling up
the lowest Landau level of the up spins while leaving the down spin
sector empty.

The charge and spin sectors are not decoupled from each other.
Firstly, the constraint of Eq.~\ref{eq:jrvb} sets the local
charge density to be the same as the local spin density. The wave
functions of the allowed states have to satisfy this local property.
Secondly, if the system is $SU(2)$ invariant, all of the states in a
given $SU(2)$ multiplet must have the same filling fraction. Since the
fully polarized states have to span all of the Jain states for a
single-layer system, we must conclude that the $SU(2)$ states which are
not in a main Jain hierarchy cannot achieve the maximum polarization.
In other terms, there is an upper bound for the spin polarization and,
hence, for the total spin itself. This argument indicates that
the filling fraction $\nu$ and
the spin $S$ of the state cannot be completely independent from each
other for the allowed states. In other words, there should exist a set
of selection rules which determines the possible combinations of
total spin and filling fraction.
Similarly, it should be possible to construct a unified theory of
all the FHQE states with $SU(2)$ symmetries, instead of the apparently
separate descriptions of the spin singlet and the fully polarizable
states that we discuss here.

We conclude this Section with a comparison of the states that are
obtained by this $SU(2)$-symmetric approach and the $U(1) \otimes U(1)$
theory. A direct inspection of the
allowed fractions, Eq.~\ref{eq:fillfrac} and Eq.~\ref{eq:allowed},
for the $U(1) \otimes U(1)$
and $SU(2)$ theories respectively, shows that they do not yield the same
allowed fractions. For instance, the ``Fermi Liquid" (compressible)
states, with the same occupancy of the two layers, allowed by
Eq.~\ref{eq:fillfrac} have filling fractions
${\frac{2}{r}}=2,1,{\frac{2}{3}},
{\frac{1}{2}}, {\frac{2}{5}}, {\frac{1}{3}}, {\frac{2}{7}},
{\frac{1}{4}}, {\frac{2}{9}}, \ldots$ ($r=1,2, \ldots$). In contrast,
the allowed $SU(2)$ ``Fermi Liquid" (compressible) states are
${\frac{2}{4s \pm 1}}=2,{\frac{2}{5}}, {\frac{2}{3}}, {\frac{2}{9}},
{\frac{2}{7}},\ldots$. Clearly, the fractions ${\frac{1}{k}}$ (with
$k=1, 2,3,\ldots$) cannot be realized as $SU(2)$ compressible states.
Among the states which appear in both hierarchies, we find an
incompressible spin unpolarized state at filling fraction
${\frac{4}{11}}$. It is worth noting that there is experimental
evidence of an incompressible state at ${\frac{4}{11}}$. It is not
possible to construct a fully polarized Jain state with this filling
fraction for single-layers although it may be constructed as a
hierarchical state. It is strange that this is the only observed
fraction for which a hierarchical construction is needed. It is
believed that the experimentally observed state is polarized.
We have checked that all of the states in levels $1$ and $2$ in the
$SU(2)$ hierarchy span the entire level $1$ $U(1) \otimes U(1)$ states.
Similarly, the level $4$ $SU(2)$ states span the level $2$  $U(1)
\otimes U(1)$ states, and the levels $3$ and $6$ $SU(2)$ states span the
level $3$ $U(1) \otimes U(1)$ states. However,  a large
number of incompressible $U(1) \otimes U(1)$
states cannot be realized as $SU(2)$ states. In a way, this may not be
surprising since the $U(1) \otimes U(1)$ symmetric theory may only
generate $SU(2)$ as a dynamical symmetry. Nevertheless, it is still
surprising since in this discussion the form of the interaction terms
has not entered and, hence, the symmetries of the Hamiltonian have not
had a chance to play
any role yet. However, more surprising is the fact that not all of the
$SU(2)$ states can be realized as $U(1) \otimes U(1)$ states. It is easy
to check, for instance, that the allowed $SU(2)$ state with
$\nu={\frac{10}{7}}$ and $S_z=0$ has no counterpart in the $U(1)
\otimes U(1)$ states (unless the condition of $S_z=0$ is relaxed and
polarized states are also considered).

\section{Ground state wave functions}
\label{sec:wave}

In this section we will make the connection between the fermion Chern-Simons
field theory and the many-body wave-function approach. We will give a
description in terms of the bilayer system since it already includes the
single layer system.

The first step that we need in order to make this connection explicit is
to find the exact asymptotic form of the density-density correlation
functions.
We need to show first that the long wavelength form of
$K_{00}^{\alpha\beta}$,
found at this
semiclassical level, saturates the $f$-sum rule. This result implies
that the non-gaussian corrections do not contribute at very small
momentum. We will use this
result to show that the absolute value squared of the ground state
wave function of this state has the Halperin \cite{halperin1} form at very long
distances, in the thermodynamic limit.

The $f$-sum rule states that the retarded density-density correlation
function $D^{\alpha\beta }_{00}$ satisfies
\begin{equation}
\int _{-\infty}^{\infty } \; {d\omega \over 2\pi} \; i \omega
       D^{\alpha\beta }_{00}(\omega,{\vec Q}) = {{\bar \rho}^{\alpha}
\over M}{\vec Q}^2  {\delta}^{\alpha\beta }
\label{eq:sum}
\end{equation}
This equation implies the conservation of the current in each
layer separately. It is easy to show that, in the basis of the total and
relative currents, {\it i.~e.\/}, in the ${\bar \rho }_{\pm} =
{\bar \rho }^{1} {\pm}{\bar \rho }^{2}$ basis, Eq.~ \ref{eq:sum} states the
conservation of ${\bar \rho }_{+}$ and ${\bar \rho }_{-}$ independently.
On the other hand, from the general definition of the polarization
tensor $K^{\alpha\beta }_{00}$, it follows that
(see for instance reference \cite{book})
\begin{equation}
K^{\alpha\beta }_{00}(x,y) = - D^{\alpha\beta }_{00}(x,y)
\label{eq:rel}
\end{equation}
Thus, Eq.~\ref{eq:sum} also holds if we replace
$D^{\alpha\beta }_{\mu\nu}$ by $K^{\alpha\beta }_{00}$ (except for an
extra minus sign).

\subsection{Ground state wave function for ($m,m,n$) states}
\label{subsec:waveunmedio}

We have found that for an ($m,m,n$) state, the leading order term in
${\vec Q}^2$ of the zero-zero component of the electromagnetic response is
given by
\begin{eqnarray}
K_{00}^{\alpha\beta } = &&- {{\bar \rho}\over 4M}\;
{{\vec Q}^2 \over {{\omega }^2 -{\omega }^2_c}
            + i \epsilon}  \left( \begin{array}{cc} 1& \;  1\\ 1& \; 1
                   \end{array}\; \right)  \nonumber \\
         && - {{\bar \rho}\over 4M}\;
{{\vec Q}^2 \over {{\omega }^2 -({(m-n) \bar \omega })^2}
            + i \epsilon}  \left( \begin{array}{cc} 1& \;  -1\\ -1& \; 1
                   \end{array}\; \right)
\label{eq:ksum}
\end{eqnarray}
Notice that the poles whose residues in $K_{00}^{\alpha\beta }$ are
proportional to ${\vec Q}^2$, have zero momentum frequency given by
$(m+n){\bar \omega}={\omega }_{c}$ and $(m-n){\bar \omega}$. It is
straightforward to check that Eq.~\ref{eq:ksum} saturates the $f$-sum
rule.
Therefore, the fermionic Chern-Simons approach gives the correct
leading order form for the density correlation function,
consistent with the $f$-sum rule, at the semiclassical level of the
approximation.

It is important to emphasize that the coefficient of the leading order
term of
$K_{00}^{\alpha\beta }$ can not be renormalized by higher order terms
in the
gradient expansion, nor in the semiclassical expansion. In the case
of the gradient expansion, it is clear that higher order terms have higher
order powers of ${\vec Q}^2$, and then, do not modify the leading order
term.
In the case of the corrections to $K_{00}^{\alpha\beta }$ originating in
higher order terms
in the semiclassical expansion, they also come with higher order powers of
${\vec Q}^2$. The reason for that is the gauge invariance of the
system which implies that higher order correlation functions must be
transverse in real space or, equivalently, have higher order powers of
${\vec Q}^2$ in momentum space.
Terms in this expansion higher than ${\vec Q}^2$ cannot change
the leading order term.

We will now follow the methods of references \cite{lopez3,lopez2} to
write the ground state wave function in the density representation.
We begin by recalling that the absolute value squared of the ground
state wave function in the density representation $|\Psi_0[\rho]|^2$ is
given by \cite{wavefnts}
\begin{eqnarray}
 |\Psi_0[{\rho}_{1},{\rho}_{2}]|^2  =&& \nonumber\\
&& \!\!\!\!\!\!\!\!\!\!\!\!\!\!\!\!\!\!\!\!\!\!\!\!\!\!
\int {\cal D} A_0^{\alpha} \;
 e^{-i\int d^{2}x \; A_0^{\alpha}({\vec x})\;\rho^{\alpha}({\vec x})}
 \lim_{A_{0}^{\alpha}(x)\to A_0^{\alpha}({\vec x})\delta (x_0)}
\langle 0|{T  e^{i\int d^3x\;A_{0}^{\alpha}(x)\;{\hat j}_0^{\alpha}(x)}}|0
\rangle
\nonumber\\
&&
\label{eq:wf1}
\end{eqnarray}
where ${\hat j}_0^{\alpha}(x)\equiv{\hat \rho}^{\alpha}(x)$.

The operators in this
expression are Heisenberg operators of the system in the absence of sources.
The vacuum expectation value in the integrand of Eq.~\ref{eq:wf1} can be
calculated from the generating functional of density correlation
functions, ${\cal Z}[{\tilde A}_{\mu}^{\alpha}]$, defined by
\begin{equation}
 \lim_{A_{0}^{\alpha} (x)\to A_{0}^{\alpha} ({\vec x})\delta (x_0)} \;
{\cal Z}[A_{0}^{\alpha} ,{\vec A}^{\alpha} =0] =
  \int {{\cal D}{\psi^{*}}} {\cal D} \psi {\cal D} a_{\mu}^{\alpha} {\ }
  e^{i S(\psi^{*},\psi ,a_{\mu}^{\alpha} , A_{\mu}^{\alpha} ) }
\label{eq:wf3}
\end{equation}
Hence, $|\Psi_0[{\rho}_{1},{\rho}_{2}]|^2$ is determined by
the generating functional of equal-\-time density correlation functions.

The path integral on the r.~h.~s.~ of Eq.~\ref{eq:wf3} can be written in
terms of the effective action $S_{eff}(A_{\mu}^{\alpha})$ for the external
electromagnetic field.
We have seen that, in the thermodynamic limit, and for weak fields, the
effective action admits the expansion given by  Eq.~\ref{eq:ems}.
Since we need only the density correlation functions, it suffice to know
the zero-zero component of $K_{\mu \nu}^{\alpha\beta}$.
In momentum space, and in the small ${\vec Q}^2$ limit,
$K_{00}^{\alpha\beta}$
is given by Eq.~\ref{eq:ksum}.
 We can see that the dominant
term in $K_{00}^{\alpha\beta}$ is of order $1/B$. Higher order terms in the
gradient expansion will contribute with higher powers of $1/B$. The same
observation applies for all the  corrections to $K_{00}^{\alpha\beta}$
originating in higher
order terms in the semiclassical expansion. Here the thermodynamic
limit is crucial since we are only taking into account fluctuations with
wavelengths short compared with the linear size of the system. The
higher order terms, which vanish like powers of ${\vec Q}^2/B$, can only
be neglected for an infinite system.

Using the results of Sections~ \ref{subsec:CS} and \ref{subsec:colle},
Eq.~\ref{eq:wf3} becomes
\begin{equation}
 \lim_{A_{0}^{\alpha}(x)\to A_{0}^{\alpha}({\vec x})\delta (x_0)} \;
{\cal Z}[A_{0}^{\alpha},{\vec A}^{\alpha}=0]=
 e^{ {i\over 2} \int d^{2}x  d^{2}y A_{0}^{\alpha}({\vec x})
(\lim_{{x_0}\rightarrow {y_0}} K_{00}^{\alpha\beta}(x,y))
A_{0}^{\beta}({\vec y}) }
\label{eq:wf4}
\end{equation}
or, by Fourier transforming the exponent, we get
\begin{equation}
 \lim_{A_{0}^{\alpha}(x)\to A_{0}^{\alpha}({\vec x})\delta (x_0)} \;
{\cal Z}[A_{0}^{\alpha},{\vec A}^{\alpha}=0]=
e^{ {i\over 2} \int {d^{2}Q\over (2\pi)^2}{\ } A_{0}^{\alpha}({\vec Q})
        ({\int\limits_{-\infty}^{\infty } {d\omega \over{2\pi}}{\ }
  K_{00}^{\alpha\beta}(\omega, \vec Q)}){\ } A_{0}^{\beta}(-{\vec Q}) }
\label{eq:wf5}
\end{equation}

The terms dropped in the exponent of Eq.~\ref{eq:wf4} and
Eq.~\ref{eq:wf5} represent equal-\-time density  correlation functions
with more than two densities. These terms give rise to three-\-body
corrections terms ( and higher) to the wave function that modify the
Jastrow form. These higher order terms, as we noted above, are of order
higher than ${\vec Q}^2$ and can be neglected.

Replacing the expression for $K_{00}^{\alpha\beta}(\omega, \vec Q)$
given by Eq.~\ref{eq:ksum} into Eq.~\ref{eq:wf5} and integrating out
$A_{0}^{\alpha}$, we obtain the following form for the absolute value
squared of the wave function
\begin{eqnarray}
|\Psi ({\vec x}_1,...,{\vec x}_{N_{1}},
&& \!\!\!\!\!\!\!\!\!\! {\vec y}_1,...,{\vec x}_{N_{2}})|^2
=
\nonumber\\
          && {\prod_{i<j=1}^{N_{1}}}{\ }|{\vec x}_i -{\vec x}_j|^{2m} \;
           {\prod_{i<j=1}^{N_{2}}}{\ }|{\vec y}_i -{\vec y}_j|^{2m} \;
     {\prod_{i=1}^{N_{1}}\prod_{j=1}^{N_{2}}}{\ }|{\vec x}_i -{\vec y}_j|^{2n}
\nonumber\\
&&{\times}\; {\exp }\big\{ -  {B\over 2} ({\sum_{i=1}^{N_{1}}}|{\vec x}_i|^2
                            + {\sum_{i=1}^{N_{2}}}|{\vec y}_i|^2 )\big\}
\label{eq:wf6}
\end{eqnarray}
where the coordinates ${\vec x}_i $ are in plane $1$, ${\vec y}_i $ are in
plane $2$, and $N_{1}=N_{2}= {N\over 2}$ for this state.
Here we have used that the eigenvalues of the local density
operator, in a Hilbert space with $N_{\alpha}$ particles, are
$\rho^{\alpha} (\vec x) =
{\sum_{i=1}^{N_{\alpha}}} {\delta (\vec x -{\vec x}_i)} - {\bar
\rho}^{\alpha}$.

The wave function of Eq.~\ref{eq:wf6} is the absolute value
squared of the Halperin wave function \cite{halperin1}. Numerical calculations
have established \cite{num} that this wave function accurately describes the
ground state wave function for the ($3,3,1$) state for $d=1.5 \ell_{c}$,
where $\ell_{c}$ is the cyclotron length.
We have shown that Eq.~\ref{eq:wf6} gives the exact form of the
ground state wave function at long distances and in the thermodynamic limit.
Since the {\it leading order term} of $K_{00}^{\alpha\beta}$
saturates the $f$-sum rule, higher order corrections in the expansion
cannot modify this result.
Arguments similar to the line of reasoning that led to Eq.~\ref{eq:wf6}
can be used to derive an analogous expression for
a single layer system \cite{lopez2}.

\subsection{Ground state wave function for ($m,m,m$) states}
\label{subsec:wavemm}

For these states, the leading order term in
${\vec Q}^2$ of the zero-zero component of the electromagnetic response is,
according to Eq.~\ref{eq:resveinte} and \ref{eq:resveintiuno}
or \ref{eq:restreintayuno}
\begin{eqnarray}
K_{00}^{\alpha\beta } = &&- {{\bar \rho}\over 4M}\;
{{\vec Q}^2 \over {{\omega }^2 -{\omega }^2_c}  + i \epsilon}
              \left( \begin{array}{cc} 1& \;  1\\ 1& \; 1
                   \end{array}\; \right)  \nonumber \\
         && - {{\bar \rho}\over 4M}\;
{{\vec Q}^2 \over {{\omega }^2 - v^{2}{\vec Q}^{2}
            + i \epsilon}  }
              \left( \begin{array}{cc} \;\; 1&  -1\\ -1& \;\; 1
                   \end{array}\; \right)
\label{eq:ksum2}
\end{eqnarray}

Following the same steps we can prove that
the leading order term of $K_{00}^{\alpha\beta }$ saturates the $f$-sum rule,
Eq. ~\ref{eq:sum}.  All the remarks on the exactness of this result
are also valid in this case.
Substituting Eq.~\ref{eq:ksum2} into the expression for the
 absolute value squared of the ground state
wave function (Eq.~\ref{eq:wf5}) we obtain
\begin{eqnarray}
|\Psi ({\vec x}_1,...,{\vec x}_{N_{1}},
&& \!\!\!\!\!\!\!\!\!\!{\vec y}_1,...,{\vec x}_{N_{2}}) |^2 =
\nonumber \\
 &&          {\prod_{i<j=1}^{N_{1}}}{\ }|{\vec x}_i -{\vec x}_j|^{2m} \;
           {\prod_{i<j=1}^{N_{2}}}{\ }|{\vec y}_i -{\vec y}_j|^{2m} \;
     {\prod_{i=1}^{N_{1}}\prod_{j=1}^{N_{2}}}{\ }|{\vec x}_i -{\vec y}_j|^{2m}
   \nonumber \\
&& \times  {\ }{\exp }\big\{ -  {B\over 2} ({\sum_{i=1}^{N_{1}}}|{\vec x}_i|^2
            + {\sum_{i=1}^{N_{2}}}|{\vec y}_i|^2 )\big\} \nonumber \\
 \times {\ } {\exp }
\big\{ -{m\; {v_{s}} \over {\omega _{c}}} &&\!\!\!\!\!\!\!\!\!\!\!\!\!
\left(
{\sum_{i,j=1}^{N_{1}}} {1\over |{\vec x}_i -{\vec x}_j|}+
{\sum_{i,j=1}^{N_{2}}} {1\over |{\vec y}_i -{\vec y}_j|}-
2 {\sum_{i=1}^{N_{1}}} {\sum_{j=1}^{N_{2}}}{1\over |{\vec x}_i -{\vec y}_j|}
\right) \big\}
\nonumber \\
&&
\label{eq:wf8}
\end{eqnarray}
where the coordinates ${\vec x}_i $ are in plane $1$, ${\vec y}_i $ are in
plane $2$, and $N_{1}=N_{2}= {N\over 2}$ for this state.

Notice that this ground state wave function is not exactly the same as
the $(m,m,m)$  Halperin wave function, to which the true ground state
approaches as $d \rightarrow 0$. There is an extra
factor which comes from the fact that there is a gapless mode in the spectrum
of collective excitations.
This contribution is analogous to the phonon contribution to the wave function
of superfluid He$_4$, and just as in that problem, it is essential
to obtain the correct properties for the spatial correlations of the ground
state.
This contribution  is very small at long distances compared to the cyclotron
radius, and it can not be written only in terms of coordinates in the
lowest Landau level.

\section{Conclusions}
\label{sec:conc}

In this Chapter, we have reviewed the fermionic Chern-Simons theory of
the 2DEG in a strong magnetic field in the regime of the fractional quantum 
Hall effect. We showed that, for various rational values of the filling fraction 
$\nu$, the mean field approximation of this theory describes a set of
composite fermions filling up an integer number of effective Landau
levels, in agreement with Jain's picture. This resulting ground state is non 
degenerate and supports a gap to all of its excitations, a property that shares 
with the true ground state of the system. By using standard techniques of
perturbation theory one can go beyond the mean field approximation in a systematic
fashion. 

The theory that we reviewed here is based on a second-quantized fermion path integral
approach. We showed that the
problem of interacting electrons moving on a plane in the presence of an
external magnetic field is equivalent to a family of systems of fermions
bound to an even number of fluxes, and that this theory has the fermions
coupled to a Chern-Simons gauge field with Chern-Simons coupling
constant $\theta =(1/ 2 \pi \times 2 s)$. The semiclassical approximation of
this system has solutions which describe incompressible liquid states,
Wigner crystals and soliton-like defects.
We worked around the liquid-like solution. This mean field solution,
or Average Field Approximation (AFA), is seen
to violate Galilean invariance explicitly. At this level of the approximation,
the center of mass of the system executes a cyclotron-like motion at the
effective cyclotron frequency. In this sense the gaussian (or semiclassical)
fluctuations are essential to restore the original symmetries of the problem.
We saw that, order-by-order in the semiclassical expansion, the
response functions obey the correct symmetry properties required by
Galilean and Gauge invariance, and by the incompressibility of the fluid.
We showed that, already at the semiclassical or gaussian level, the
low-momentum limit of the density correlation function saturates the
{\it f}-sum rule, and in that sense this result is exact, {\it i.~e.\/}, it can
not be renormalized by higher order corrections.
We showed how our methods can be used to calculate the density-current
correlation functions and, from them, to extract the Hall conductance.
We found that it has the correct value already at the gaussian
level of the approximation. We also  obtained the spectrum of
collective excitations in the low-momentum
limit for short-range and for Coulomb interparticle pair potentials.

We have also presented a generalization of our approach to the study of
the FQHE in double-layer
systems and for unpolarized and partially polarized single-layer systems.
These systems have a richer class of behaviors and excitation
spectra. We have used the fermionic Chern-Simons theory to
investigate the physics of these states, including results on their
response functions and for the quantum numbers of their topological
excitations. We have also given a specific realization of Haldane's
criterion for topological stability.

Finally, we gave an explicit construction of  the ground state wave
functions from our semiclassical field theory. We show that their universal
properties are a consequence of general principles, {\it i.~e.\/},
incompressibility and Galilean invariance which determine the analytic
structure of the equal-time density correlation functions at long distances.

\section*{Acknowledgments}

EF wishes to thank the Theoretical Physics Group of the Department of
Physics of Oxford University, where part of this work was written,
for its kind hospitality.
This work was supported in part by the National Science Foundation 
through the grant NSF DMR94-24511 at the University of Illinois at 
Urbana-Champaign, and by a Glasstone Research Fellowship in Sciences 
and a Wolfson Junior Research Fellowship (AL).

\end{document}